\documentclass[12pt,twoside]{article}
\usepackage{correl}

\def\TheTitle{Interplay of\newline Kondo Effect and RKKY Interaction}
\def\TheHeading{Kondo Effect and RKKY Interaction}
\def\TheAuthor{Johann Kroha}
\def\TheInstitute{Physikalisches Institut, Universit\"at Bonn}
\def\TheAddress{Nussallee 12, 53115 Bonn \newline Germany}
\def\TheChapter{5}                       

\usepackage{dsfont}   

\newcommand{\ve}[1]{\mathbf{#1}}

\newcommand{\ket}[1]{|#1 \rangle}

\newcommand{\im}{\text{Im}}

\newcommand{\phdagger}{{\phantom{\dagger}}}
\newcommand{\bsig}{\boldsymbol\sigma}        
\newcommand{\Eq}[1]{Eq.~(\ref{#1})}          
\newcommand{\Eqs}[2]{Eqs.~(\ref{#1}), (\ref{#2})}

\begin{document}
\MakeTitle           


\section{Introduction and overview}
\label{sec:intro}

Magnetic interactions in a metal involving localized magnetic moments 
give rise to a wealth of phenomena, ranging from the Kondo effect to 
magnetic ordering and quantum phase transitions. We give a brief overview 
of such phenomena, before in the main part of these lecture notes 
we will focus on a detailed description of the interplay of interactions 
that tend to quench the local moments or that tend order them.

When a magnetic ion is placed in a metallic host, the Kondo 
effect \cite{Kondo64,Hewson93} occurs: 
Conduction electrons at the Fermi level, i.e. at zero 
excitation energy, are in resonance with a flip of the two-fold 
degenerate spin ground state of the magnetic ion. As the temperature $T$ 
is lowered, the electrons become confined to the Fermi surface, so that
more and more electrons contribute to this resonant quantum spin-flip 
scattering, leading to a diverging spin scattering amplitude.  
Hence, when the spin exchange coupling $J_0$ between the localized moments 
\index{local moment}
and the itinerant conduction electrons is antiferromagnetic, a 
many-body spin-singlet state between the impurity spin and the 
conduction electron spins is formed below a characteristic temperature,
the Kondo temperature $T_K$. This, however, means that electrons that do 
not contribute to the singlet bound state, experience merely potential 
scattering rather than spin scattering, i.e., the impurity spin is effectively
removed from the system. This effect is called spin screening. 
\index{spin screening} 
The scattering rate and other physical quantities thus settle smoothly to 
constant values, leading to Fermi liquid behavior for $T\ll T_K$ 
\cite{Hewson93}. The Kondo temperature is found to be exponentially 
small in the exchange coupling, $T_K=D_0\exp [-1/(2N(0)J_0)]$, with the density 
of states at the Fermi level $N(0)$ and the conduction band width $D_0$.
The entirety of complex phenomena sketched above, involving increase 
of the spin scattering amplitude implying anomalous transport properties, 
followed by spin screening and the formation of a narrow, but smooth 
resonance of width $T_K$ in the electronic spectrum at the Fermi energy, 
comprises the Kondo effect. \index{Kondo effect}

When there are several or many localized magnetic moments in a metal, 
for instance arranged on a lattice, the same spin-exchange coupling $J_0$ 
that induces the Kondo effect, induces also a magnetic interaction between 
the localized spins:
The local moments can exchange their spins, mediated by two conduction 
electrons scattering from and traveling between the impurity sites.
Since this effective, long-range spin-exchange coupling $K$ involves 
two elementary scattering events between electron and impurity spins, 
it is of order $K\propto N(0)J_0^2$.
It can be ferro- or antiferromagnetic due to the 
long-range, spatial oscillations of the conduction electron
density correlations. This conduction-electron-mediated spin 
interaction was first considered by Ruderman and Kittel \cite{Ruderman54},
Kasuya \cite{Kasuya56} and Yosida \cite{Yosida57} and is therefore 
called RKKY interaction. \index{RKKY interaction}
The RKKY interaction usually dominates the magnetic dipole-dipole
coupling as well as the direct exchange coupling 
between neighboring local moments because of the short spatial extent of these 
couplings or the exponentially small overlap of the local moment wave 
functions on neighboring lattice sites.

\begin{figure}
\centering
\includegraphics[width=0.75\linewidth]{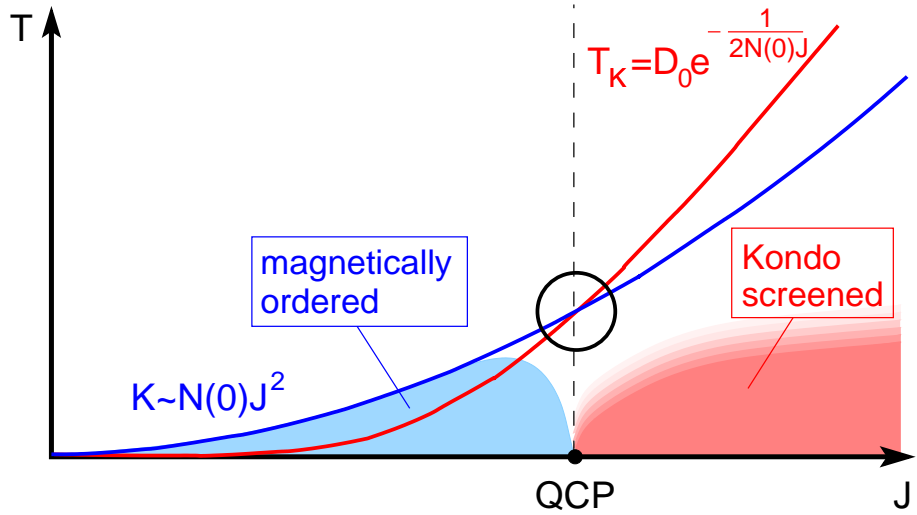}
\caption{Doniach's phenomenological phase diagram \index{Doniach phase diagram}
for the phase transition between an RKKY-induced, magnetically ordered phase 
and the Kondo screened, paramagnetic phase. 
The phase transition occurs when the RKKY coupling $K$
of a local moment to all surrounding moments becomes equal to the Kondo 
singlet binding energy $T_K$ (black circle). 
While the RKKY coupling is $K\sim N(0)J_0^2$,
the Kondo energy $T_K=D_0\exp [-1/(2N(0)J_0)]$ is exponentially small in
the bare, local spin exchange coupling $J$. 
Therefore, the RKKY coupling always dominates for small values of $J_0$.   
\label{fig:doniach}
}
\end{figure}

In a Kondo lattice, the local Kondo coupling and the 
RKKY interaction favor different ground states. 
The Kondo coupling leads to a paramagnetic Fermi liquid state without 
local moments. In this state, 
the local orbitals, whose spectrum has a Kondo resonance at the 
Fermi energy, hybridize with each other and eventually become lattice 
coherent at low temperatures to form Bloch-like quasiparticle states. 
As a result, a narrow band crossing the Fermi energy is formed. 
Its bandwidth is controlled by the Kondo resonance width $T_K$. It thus
gives rise to an exponentially strong effective mass enhancement of 
roughly $m^*/m\approx \exp [1/(2N(0)J_0)]$, which lends the name 
``heavy Fermi liquid'' to this state \cite{Loehneysen07}.\\
By contrast, the RKKY interaction tends to induce magnetic order
of the local moments. It was pointed out early on by Doniach  
\cite{Doniach77} that, therefore, the Kondo spin screening of 
the local moments should eventually break down \index{Kondo breakdown}
and give way to magnetic order,
when the RKKY coupling energy becomes larger than the characteristic energy 
scale for Kondo singlet formation, the Kondo temperature $T_K$, 
see Fig.~\ref{fig:doniach}. Thus, one expects a  $T=0$ quantum phase transition 
(QPT) to occur \cite{Loehneysen07}, with the local spin exchange 
coupling $J_0$ as the control parameter. \index{quantum phase transition}
If and how the Kondo breakdown occurs at a magnetic QPT is, 
however, controversial. In fact, several QPT scenarios in heavy-fermion 
systems are conceivable.\\ 
(1) The heavy Fermi liquid, like any other Fermi liquid, may undergo a 
spin-density wave (SDW) instability, leading to critical fluctuations of the 
bosonic magnetic order parameter but leaving the fermionic, heavy 
quasiparticles intact. This scenario is well described by the pioneering 
works of Hertz, Moriya and Millis \cite{Hertz76,Moriya85,Millis93}.\\ 
(2) The local fluctuations of the magnetization, coupling to the nearly 
localized, heavy quasiparticles, may become critical (divergent) and 
thereby destroy the heavy Fermi liquid (local quantum criticality)
\cite{Si01,Coleman01}.\\

(3) At the phase transition the Kondo effect and, hence, the heavy-fermion 
band vanish, which leads to an abrupt change of the Fermi surface 
(Fermi volume collapse). It has been proposed \cite{Senthil04} that the 
Fermi surface fluctuations associated with this change may selfconsistently 
destroy the Kondo singlet state.\\
(4) Most recently, a scenario of critical quasiparticles has been put forward, 
characterized by a diverging effective mass and a singular quasiparticle 
interaction which is selfconsistently generated by the nonlocal 
order-parameter fluctuations of an impending SDW instability 
\cite{Woelfle11,Woelfle14, Woelfle16}. 
Intriguing in its generality and similar in spirit to Landau's Fermi liquid 
theory, this scenario does, however, not invoke Kondo physics and, 
thus, does not address the specific problems associated with the 
Kondo destruction like Fermi volume collapse or the possibility of small, 
localized magnetic moments in the magnetically ordered phase.

While the Hertz-Millis-Moriya scenario (1) is described by a 
critical field theory of the bosonic, magnetic order parameter alone,
the complete understanding of the breakdown scenarios (2), (3) and (4) would 
require a field theory for the fermionic degrees of freedom forming the 
Kondo effect and the heavy quasiparticles, coupled to the bosonic order 
parameter field. In lack of such a complete theory, 
these scenarios presume that specific fluctuations, (2) local 
fluctuations, (3) Fermi surface fluctuations or 
(4) antiferromagnetic fluctuations, become soft for certain values of the
system parameters and, thus, dominate the QPT. Therefore, the 
conditions for these scenarios to be realized are controversial. 

In these lecture notes we consider the interplay of Kondo screening and RKKY 
interaction within the Kondo lattice model. 
We derive the phenomena of the single-impurity Kondo model 
in section \ref{sec:Kondo}, thereby introducing important concepts and 
techniques, like the fermionic representation spin, universality and 
the analytic (perturbative) renormalization group. 
Section \ref{sec:RKKY} presents the oscillatory RKKY coupling, calculated 
as a second-order spin exchange process, mediated by the conduction electrons.  
In section \ref{sec:Kondo-RKKY} we show how the Kondo singlet formation as
well as the RKKY interaction can be incorporated on the same footing in an
analytic renormalization group treatment. It predicts a universal 
Kondo destruction as function of the RKKY coupling parameter. 
We conclude in section \ref{sec:conclusion} with a discussion 
how this theory may set the stage for a more complete quantum field theory 
of heavy-fermion QPTs with Kondo breakdown.


\section{Kondo effect and renormalization group}
\label{sec:Kondo}

In this section we recollect the essential physics of a single Kondo  
impurity in a metal \index{Kondo effect} 
and provide the calculational tools for their derivation.  
We consider the single-impurity Kondo model, 
\begin{equation}
H = \sum_{\mathbf{k},\sigma} 
\varepsilon_{\mathbf{k}} \; c_{\mathbf{k} \sigma}^\dagger c_{\mathbf{k} \sigma}^\phdagger +
J_0\, \hat{\mathbf{S}} \cdot \hat{\mathbf{s}} 
\label{eq:KondoHamiltonian}
\end{equation}
where $c_{\mathbf{k} \sigma}^{\phdagger}$, $c_{\mathbf{k}\sigma }^{\dagger}$
denote the conduction 
($c$-) electron operators with momentum $\mathbf{k}$ and 
dispersion $\varepsilon_{\mathbf{k}}$. $\hat{\mathbf{S}}$ is the impurity spin 
operator at site $\mathbf{x}=0$ which is locally coupled to the spins 
of the conduction electrons on that site, $\hat{\mathbf{s}}$, via a Heisenberg 
exchange coupling $J_0$. We have
\begin{equation}
\hat{\mathbf{s}} = \sum_{\mathbf{k},\mathbf{k}',\,\sigma , \sigma '}
c_{\mathbf{k} \sigma}^{\dagger}\, \bsig_{\sigma \sigma '}\, c_{\mathbf{k}' \sigma '}^{\phdagger},
\label{eq:cspin}
\end{equation}
with $\bsig =(\sigma_x,\,\sigma_y,\,\sigma_z)^T$ the vector of Pauli matrices,
\begin{eqnarray}
\sigma_x=\begin{pmatrix}
0 & 1\\
1 & 0\\
\end{pmatrix}
\qquad
\sigma_y=\begin{pmatrix}
0 & -i\\
i & 0\\
\end{pmatrix}
\qquad
\sigma_z=\begin{pmatrix}
1 &  0\\
0 & -1\\
\end{pmatrix}.
\end{eqnarray}
In \Eq{eq:cspin} the conduction spin eigenvalue $1/2$ has been 
absorbed in the coupling constant $J_0$, by convention, 
and we use units $\hbar =1$ throughout. 
The local spins $\hat{\mathbf S}$ will henceforth be termed $f$-spins, 
as they are typically realized in heavy fermion systems by the rare-earth $4f$ 
electrons.

\subsection{Pseudofermion representation of spin}
\label{subsec:pseudofermion} 

A field theoretical treatment, like the standard functional integral or
Wick's theorem and many-body perturbation theory, requires that the 
corresponding field operators obey canonical commutation rules, i.e.,
their (anti)commutators must be proportional to the unit operator.
However, the spin operators $\hat{\mathbf{S}}$ obey the SU(2) algebra. 
In order to overcome this difficulty, we use the fermionic representation 
of spin, first introduced by Abrikosov \cite{Abrikosov65}. 
For each of the basis states spanning the impurity spin Hilbert space,  
$\ket{\sigma}$, $\sigma=\uparrow,\,\downarrow$, fermionic creation and 
destruction operators $f_{\sigma}^{\dagger}$, $f_{\sigma}^{\phdagger}$ are
introduced according to 
$\ket{\sigma}=f_{\sigma}^{\dagger}\ket{vac}$, where $\ket{vac}$ denotes the 
vacuum state (no impurity spin present). The impurity spin operator 
$\mathbf{S}$ then reads, \index{pseudofermion representation of spin}
\begin{equation}
\hat{\mathbf{S}} = \frac{1}{2}\sum_{\tau , \tau '}
f_{\tau}^\dagger\, \bsig_{\tau \tau '}\, f_{\tau '}^{\phdagger}.
\label{eq:fspin}
\end{equation}
That is, the operator on the right-hand side and $\hat{\mathbf{S}}$ have   
identical matrix elements in the physical spin Hilbert space. 
However, repeated action of the fermionic operators would permit 
unphysical double occupancy or no occupancy 
of the spin states $\ket{\uparrow}$, $\ket{\downarrow}$. The dynamics are 
restricted to the physical spin space by imposing the operator constraint
\begin{equation}
\hat Q=\sum_{\tau} f_{i \tau}^\dagger f_{i\tau}^\phdagger = \mathds{1}.
\label{eq:constraint}
\end{equation}
\Eqs{eq:fspin}{eq:constraint} constitute the exact pseudofermion 
representation of the spin $s=1/2$.

The impurity-spin operator and, hence, the equation of motion with the 
Hamiltonian (\ref{eq:KondoHamiltonian}) are symmetric under the local
U(1) gauge transformation \index{gauge symmetry}
\begin{equation}
f_{\tau}\rightarrow {\rm e}^{-i\phi (t)} f_{\tau} , \qquad
i\frac{d}{dt} \rightarrow  i\frac{d}{dt} -\frac{\partial \phi(t)}{\partial t}, 
\label{eq:gauge}
\end{equation}
with an arbitrary, time-dependent phase $\phi (t)$. It is closely related to
the conservation of the pseudofermion number $\hat{Q}$. 

\textit{Projection onto the physical Hilbert space.} 
The exact projection of the dynamics onto the physical sector of Fock 
space with $Q=1$, is performed by the following procedure. 
Consider first the grand canonical ensemble with respect to $Q$, 
defined by the statistical operator
\begin{equation}
\hat \rho _G = \frac{1}{Z_G} {\rm e}^{-\beta (H+\lambda Q)},
\label{eq:stat_op}
\end{equation}
where $Z_G={\rm tr}[{\rm exp}\{-\beta (\hat H+\lambda \hat Q)\}]$ 
is the grand canonical partition function, $-\lambda$ 
the associated chemical potential and $\beta=1/k_BT$ 
the inverse temperature. The trace extends over the complete Fock space,
including summation over $Q=0,\,1,\,2$. The grand canonical expectation 
value of an observable $\hat A$ acting on the impurity spin space is 
defined as
\begin{equation}
\langle \hat A\rangle _G (\lambda)= {\rm tr} [\hat \rho _G \hat A] .
\label{eq:exp_val}
\end{equation}
The physical expectation value of $\hat A$, $\langle \hat A\rangle$, 
must be evaluated in the canonical ensemble with fixed $Q=1$. It can be
obtained from the grand canonical expectation value as 
\cite{Abrikosov65},
\begin{equation}
\langle \hat A\rangle :=
\frac { \mbox{tr}_{Q=1}\bigl[\hat A e^{-\beta \hat H} \bigr]} 
      { \mbox{tr}_{Q=1}\bigl[e^{-\beta \hat H} \bigr]} =
\lim _{\lambda \rightarrow \infty}
\frac { \mbox{tr}\bigl[\hat A e^{-\beta [\hat H+\lambda (\hat Q-1)]} \bigr]} 
      { \mbox{tr}\bigl[ \hat{Q} e^{-\beta [\hat H+\lambda (\hat Q-1)]} \bigr]} =
\lim _{\lambda \rightarrow \infty} 
\frac {\langle \hat A \rangle _{G}(\lambda )}{\langle \hat Q \rangle _{G}(\lambda )}
\label{eq:projection}  
\end{equation}
Here, all terms of the grand canonical traces in the numerator and in the 
denominator with $Q>1$ are projected away by the limit $\lambda\to\infty$. 
In the denominator, the operator $\hat{Q}$ makes all terms with $Q=0$ vanish.
In the numerator, the observable $\hat{A}$ acts on the impurity spin 
space and hence is a power of $\hat{S}$, Eq.~\ref{eq:fspin}, which 
vanishes in the $Q=0$ subspace. Therefore, in the numerator and in the 
denominator precisely the canonical traces over the physical sector $Q=1$
remain, as required. It follows that any impurity-spin correlation function
can be evaluated as a pseudofermion correlation function in the 
unrestricted Fock space, where Wick's theorem and the decomposition in terms 
of Feynman diagrams with pseudofermion propagators are valid, and taking the
limit $\lambda\to\infty$ at the end of the calculation. Note that for the 
$c$ electron spin, \Eq{eq:cspin}, the $Q=1$ projection is not needed,
because for the noninteracting $c$-electrons doubly occupied or empty 
states are allowed. 

\textit{Diagrammatic rules.}  
We will now show that the limit $\lambda\to\infty$ translates into 
simple diagrammatic rules for the evaluation of impurity Green's and 
correlation functions. 
We denote the local $c$ electron Green's function at the impurity site by
$G_{c\sigma}(i\omega_n)$ and the bare, grand canonical pseudofermion Green's 
function by $G_{f\sigma}^G(i\omega_n)$,
\begin{eqnarray}  
G_{c\sigma}(i\omega_n)&=&\sum_{\mathbf{k}}\,
\frac{1}{i\omega_n-\varepsilon_{\mathbf{k}}}
\label{eq:Gc}\\
G_{f\sigma}^G(i\omega_n)&=&\frac{1}{i\omega_n-\lambda},
\label{eq:Gf}
\end{eqnarray}  
with the fermionic Matsubara frequencies $\omega_n=\frac{\pi}{\beta}(2n+1)$. 
Consider first $\lim_{\lambda\to 0}\langle\hat{Q}\rangle_G(\lambda)$. 
Using standard, complex contour integration, we obtain 
\begin{eqnarray}
\langle\hat{Q}\rangle_G(\lambda) &=& 
\sum_{\sigma} \frac{1}{\beta} \sum_{\omega_n} G_{f\sigma}^G(i\omega_n) =
-\sum_{\sigma} \oint \frac{dz}{2\pi i}\, f(z)\, 
G_{f\sigma}^{G}(z) \nonumber\\
&=&
-\sum_{\sigma} \int_{-\infty}^{+\infty} \frac{d\varepsilon}{2\pi i}\,f(\varepsilon)\,
\left[ G_{f\sigma}^{G}(\varepsilon +i0) - G_{f\sigma}^{G}(\varepsilon -i0) \right],
\label{eq:contourint1}
\end{eqnarray}
where $f(\varepsilon)=1/(e^{\beta\varepsilon}+1)$ is the Fermi function, and
the $\varepsilon$-integral extends along the branch cut 
of $G_{f\sigma}^{G}(z)$ at the real frequency axis, $\im z=0$.  
We can now perform a specific gauge transformation of the operators,
$f_{\tau}\to {\rm e}^{-i\lambda t}f_{\tau}$. It implies, by virtue of
\Eq{eq:gauge}, a shift of all pseudofermion energies in a 
diagram by $\varepsilon\to\varepsilon +\lambda$. It
eliminates $\lambda$ from the pseudofermion propagator and casts it into 
the argument of the Fermi function. Thus, we have
\begin{eqnarray}
\langle\hat{Q}\rangle_G(\lambda) &=& 
-\sum_{\sigma} \int_{-\infty}^{+\infty} \frac{d\varepsilon}{\pi }\, 
f(\varepsilon+\lambda)\,\im G_{f\sigma}(\varepsilon +i0)  \nonumber \\
&\stackrel{\lambda\to\infty}{\longrightarrow}&
{\rm e}^{-\beta\lambda}\sum_{\sigma} 
\int_{-\infty}^{+\infty} \frac{d\varepsilon}{\pi }\, 
{\rm e}^{-\beta\varepsilon}\,\im G_{f\sigma}(\varepsilon +i0),
\label{eq:contourint2} 
\end{eqnarray}
where $G_{f\sigma}(\varepsilon +i0)\equiv 
G_{f\sigma}^G(\varepsilon+\lambda +i0)=1/(\varepsilon+i0)$
is independent of $\lambda$. 

The result \Eq{eq:contourint2} can be generalized by explicit 
calculation to arbitrary Feynman diagrams involving $f$- and $c$-Green's 
functions: 
(i) Each complex contour integral includes one distribution function 
$f(z)$. The integral can be written as the sum of integrals along the 
branch cuts at the real energy axis of all propagators appearing in the diagram.
(ii) Consider now one term of this sum. The argument of the distrubution 
function $f(\varepsilon)$ in that term is real and always equal to the 
argument $\varepsilon$ of that propagator $G$ along whose branch cut the 
integration extends. 
(iii) The above energy-shift gauge transformation applies to all 
pseudofermion energies $\omega$ in the diagram and, thus, cancels the 
parameter $\lambda$ in all pseudofermion propagators, 
$G_{f\sigma}^G(\omega)\to G_{f\sigma}(\omega)$. 
(iv) If in the considered term the integral is along 
a pseudofermion branch cut, this gauge transformation also shifts the 
argument of the distribution function, 
$f(\varepsilon)\to f(\varepsilon+\lambda)$, by virtue of (ii), i.e., the 
the pseudofermion branch cut integral vanishes 
$\sim {\rm  e}^{-\beta\lambda}$, as in \Eq{eq:contourint2}. 
If the integral is along a $c$-electron branch cut, the argument of 
$f(\varepsilon)$ is not affected by the gauge transformation, and the 
integral does not vanish. 

This derivation can be summarized in the following diagrammatic rules
for ($Q=1$)-projected expectation values:
\begin{itemize}
\item[(1)]
In a diagrammatic part that consists of a product of 
$c$- and $f$-Greens's functions, 
only the integrals along the $c-$electron branch cuts contribute.\\[-1.0cm]
\item[(2)]
A closed pseudofermion loop contains only pseudofermion branch cut integrals
and thus carries a factor ${\rm e}^{-\beta\lambda}$.\\[-1.0cm]
\item[(3)]
Each diagram contributing to the 
projected expectation value of an impurity spin observable, 
$\langle \hat A \rangle$, contains exactly one closed pseudofermion
loop per impurity site, because the factor of ${\rm e}^{-\beta\lambda})$ 
cancels in the numerator and denominator of \Eq{eq:projection}, 
and higher order loops vanish by virtue of rule (2). 
\end{itemize}

We note in passing that the pseudofermion representation can be 
generalized in a straight-forward way to higher local spins than $S=1/2$ by 
choosing in \Eq{eq:fspin} a respective higher-dimensional representation 
of the spin matrices and defining the contraint $\hat{Q}=\mathds{1}$ as 
before, with a summation over all possible spin orientations $\tau$. 
It can alse be extended to include local charge fluctuations by means 
of the slave boson representation 
\cite{Barnes76,Coleman84,Kroha98,Kroha97}.

\subsection{Perturbation theory}
\label{subsec:PT}

It is instructive to analyze the scattering of a conduction electron from 
a spin impurity in perturbation theory, \index{perturbation theory} 
because this will visualize the physical origin of its singular behavior. 
The perturbation theory can be efficiently 
evaluated with the formalism developed in section \ref{subsec:pseudofermion}.

\begin{figure}
\centering
\includegraphics[width=0.85\linewidth]{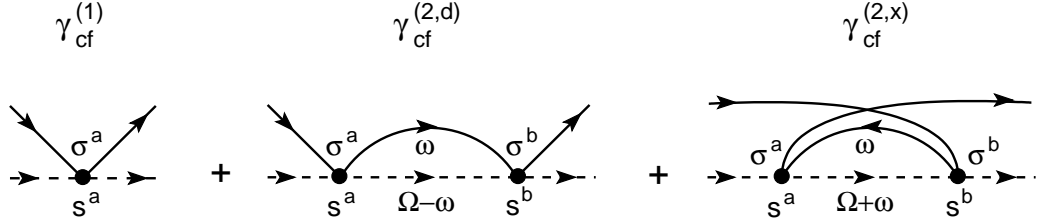}
\caption{\label{fig:PT}
Conduction electron-impurity spin vertex $\hat\gamma_{cf}$ 
of the single-impurity Kondo model up to 2nd order in the spin exchange 
coupling $J_0$. Conduction electron propagators are denoted by solid, 
pseudofermion propagators by dashed lines. 
$\hat\gamma^{2,d}$ and $\hat\gamma^{2,x}$ represent the 2nd-order direct and 
exchange terms, respectively. The external lines 
are drawn for clarity and are not part of the vertex.
}
\end{figure}

With the Kondo Hamiltonian \Eq{eq:KondoHamiltonian} the 
conduction electron-impurity spin vertex $\hat\gamma_{cf}$ can be read off
from the diagrams in Fig.~\ref{fig:PT}. 
Denoting the vector of Pauli matrices acting in $c$-electron spin space 
by $\bsig =(\sigma^x,\sigma^y,\sigma^z)^T$ and the vector of Pauli matrices 
in $f-$spin space by ${\bf s}=(s^x,s^y,s^z)^T$, $\hat\gamma_{cf}$ reads in 
first and second order of $J_0$,
\begin{eqnarray}
\hat\gamma_{cf}^{(1)}   &=& \frac{1}{2}J_0 \left( {\bf s}\cdot\bsig \right)
\label{eq:gamma1}\\
\hat\gamma_{cf}^{(2,d)} &=& -\frac{1}{4}J_0^2 
\sum_{a,b=x,y,z}\left( s^b \sigma^b \right)\left( s^a \sigma^a \right)\
\frac{1}{\beta} \sum_{\omega_n}
G_{c}(i\omega_n)G_{f}(i\Omega_m-i\omega_n)|_{\lambda\to\infty} 
\label{eq:gamma2d}\\
\hat\gamma_{cf}^{(2,x)} &=& + \frac{1}{4}J_0^2 
\sum_{a,b=x,y,z}\left( s^b \sigma^a \right)\left( s^a \sigma^b \right)\
\frac{1}{\beta} \sum_{\omega_n}
G_{c}(i\omega_n)G_{f}(i\Omega_m+i\omega_n)|_{\lambda\to\infty}, 
\label{eq:gamma2x}\end{eqnarray}
where matrix multiplications in the $f$- and $c$-spin spaces are implied,
and the sum $\sum_{a=x,y,z}$ represents the scalar product in position space. The 
relative minus sign between $\hat\gamma_{cf}^{(2,d)}$ and $\hat\gamma_{cf}^{(2,x)}$
arises because of the extra fermion loop in the exchange term 
$\hat\gamma_{cf}^{(2,x)}$. Note that the order of the Pauli matrices in 
\Eqs{eq:gamma2d}{eq:gamma2x} is crucial.
It is determined by their order along the $c$-electron or $f$-particle 
lines running through the diagram. Thus, in $\hat\gamma_{cf}^{(2,x)}$ the order 
of $c$-electron Pauli matrices is reversed with respect to  
$\hat\gamma_{cf}^{(2,d)}$.

The spin-dependent part of $\hat\gamma_{cf}^{(2,d)}$, $\hat\gamma_{cf}^{(2,x)}$ can 
be evaluated using the SU(2) spin algebra, $ \sigma^{a}\sigma^{b} = \sum_{c=x,y,z}
i \varepsilon^{abc} \sigma^{c} +\delta^{ab}\mathds{1}$ for $a,\,b=x,\,y,\,z$,
where $\mathds{1}$ is the unit operator in spin space,   
$\varepsilon^{abc}$ the totally antisymmetric unit tensor and 
$\delta^{ab}$ the Kronecker-$\delta$,
\begin{eqnarray}
{\rm d:} \qquad \sum_{a,b=x,y,z} s^bs^a\otimes \sigma^b\sigma^a &=&
-2\,  {\bf s}\cdot \bsig +3\, \mathds{1}\otimes\mathds{1}
\label{eq:spinvertex_d}\\
{\rm x:} \qquad \sum_{a,b=x,y,z} s^bs^a\otimes \sigma^a\sigma^b &=&
\phantom{-}2\, {\bf s}\cdot\bsig +3\, \mathds{1}\otimes \mathds{1}
\label{eq:spinvertex_x}
\end{eqnarray}
For scattering at the Fermi energy ($\Omega =0$), 
the energy-dependent factors in \Eqs{eq:gamma2d}{eq:gamma2x} are 
\begin{eqnarray}
{\rm d:} \qquad \frac{1}{\beta} \sum_{\omega_n} 
G_{c}(i\omega_n)G_{f}^G(-i\omega_n)|_{\lambda\to\infty}
&=& \oint \frac{dz}{2\pi i} [1-f(z)] G_{c}(z)G_{f}^G(-z)|_{\lambda\to\infty}
\nonumber\\
&=&N(0)\int_{-D_0}^{D_0} d\varepsilon\, \frac{1-f(\varepsilon)}{\varepsilon}
\label{eq:E_direct}\\
{\rm x:} \qquad \ \ \ \, \frac{1}{\beta} \sum_{\omega_n}
G_{c}(i\omega_n)G_{f}^G(i\omega_n)|_{\lambda\to\infty}
&=& - \oint \frac{dz}{2\pi i} f(z) G_{c}(z)G_{f}^G(z)|_{\lambda\to\infty}
\nonumber \\
&=& - N(0)\int_{-D_0}^{D_0} d\varepsilon\, \frac{f(\varepsilon)}{\varepsilon},
\label{eq:E_exchange}
\end{eqnarray}
where we have assumed the Fermi energy in the center of the band of 
half bandwidth $D_0$, with a flat conduction electron density of states
$N(0)=\im G_c(0-i0)/\pi$.
We see (c.f.~Fig.~\ref{fig:PT}) that in the direct term ($d$) the 
intermediate electron must scatter into an unoccupied state, 
$1-f(\varepsilon)$, while in the exchange term ($x$) the intermediate 
electron comes from an occupied state, $f(\varepsilon)$ and then leaves 
the impurity. Collecting all terms, we obtain 
$\hat\gamma_{cf}=\hat\gamma_{cf}^{(1)}+\hat\gamma_{cf}^{(2d)}+\hat\gamma_{cf}^{(2x)}$ 
as  
\begin{eqnarray}
\hat\gamma_{cf} 
&=& \frac{1}{2}J_0 \left( {\bf s}\cdot\bsig \right)\,\left[ 1 + 
N(0)J_0\int_{-D_0}^{D_0} d\varepsilon\, 
\frac{1-2f(\varepsilon)}{\varepsilon}\ + {\cal O}(J_0^2) 
\right]   \nonumber\\
&\approx&\frac{1}{2}J_0 \left( {\bf s}\cdot\bsig \right)\,
\left[ 1 + 2N(0)J_0 \ln\left(\frac{D_0}{T} \right) 
\ +{\cal O}(J_0^2) \right] 
\label{eq:PT_log}
\end{eqnarray}
The calculation clearly shows the physical origin of the logarithmic
behavior: the presence of a sharp Fermi edge in the phase space 
available for scattering, i.e., in the integrals of 
\Eqs{eq:E_direct}{eq:E_exchange}, and quantum spin-flip scattering
with the nontrivial SU(2) algebra. If the reversed order of 
Pauli matrices in the exchange term 
would not introduce a minus sign in the spin channel, \Eq{eq:spinvertex_x}, 
the logarithmic terms would cancel, like in the potential scattering channel, 
instead of adding up. 
It is also important that the impurity is localized, because otherwise 
an integral over the exchanged momentum (recoil) would smear the logarithmic 
singularity.

\Eq{eq:PT_log} exhibits a logarithmic divergence for low temperatures $T$.
\index{perturbation theory, singular}
It signals a breakdown of perturbation theory when the 2nd-order 
contribution to $\hat\gamma_{cf}$ becomes equal to the 1st-order contribution.
This happens at a characteristic temperature scale, which can be read
off from \Eq{eq:PT_log}, the Kondo temperature
\begin{equation}
T_K=D_0\,{\rm e}^{-1/(2N(0)J_0)}.
\label{eq:TK}
\end{equation}
Below $T_K$ perturbative calculations about the weak-coupling state 
break down. To describe the complex physics outlined in the introduction,
more sophisticated techniques, predominantly numerical or exact solution
methods, are required. The logarithmic behavior of the perturbation 
expansion, however, sets the stage for the development of the 
renormalization group method, to be developed in the next section, and
which is particularly useful for analytically studying the interplay of 
Kondo screening and RKKY interaction.

\subsection{Renormalization group}
\label{subsec:RG}

\begin{figure}
\centering
\includegraphics[width=0.95\linewidth]{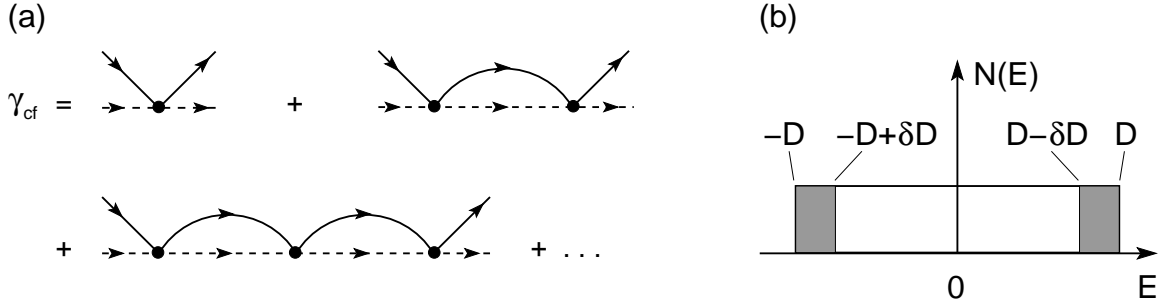}
\caption{\label{fig:RG}
Universality and perturbative renormalization group. 
(a) T-matrix resummation of the $c-f$ vertex. The sum contains, for 
each conduction electron-pseudofermion bubble (direct diagram) shown, the
exchange diagram, which is not shown for clarity.
(b) Scheme for the cutoff reduction $D\to D-\delta D$. 
}
\end{figure}

Since the logarithm is a scale invariant function, there is the 
possibility that the resummation of a logarithmic perturbation 
expansion leads to universal behavior in the sense that variables like 
energy $\omega$, temperature $T$, etc., can be expressed in units of a 
single scale, $T_K$, in such a way that all physical quantities 
are functions of the dimensionless variables, $\omega/T_K$, $T/T_K$, etc.,
only and do not explicitly depend on the microscopic parameters of the 
Hamiltonian, like $J_0$, $D_0$ and $N(0)$. 
For the Kondo model, this extremely remarkable property 
can be visualized by a T-matrix-like, partial resummation of the $c-f$ 
vertex, as sketched in Fig.~\ref{fig:RG}~(a). The resummation results 
in a geometric series for the total $c-f$ vertex or the effective
coupling constant $\tilde J$, 
\begin{eqnarray}
N(0)\tilde J  &=& 2N(0)J_0 
\left[ 1 + 2N(0)J_0 \ln\left(\frac{D_0}{T} \right) 
\ + \left( 2N(0)J_0 \ln\left(\frac{D_0}{T} \right) \right)^2 + \dots \right]\\ 
&=& \frac{2N(0)J_0}{1-2N(0)J_0 \ln\left(\frac{D_0}{T}\right)} =
\frac{1}{\ln\left(\frac{T}{T_K}\right)}, 
\end{eqnarray}
which converges for $T>T_K$. It is seen that, as a consequence of the
logarithmic behavior, in the last expression 
the microscopic parameters $J_0$, $D_0$ and $N(0)$ indeed
conspire to form the Kondo temperature $T_K$ of \Eq{eq:TK} as the only scale
in the problem. This universal behavior is inherited by physical quantities, 
like relaxation rates, transport properties, etc., since they can be expressed 
in terms of the total $c-f$ vertex.     
Although the above is only a heuristic argument and other contributions, 
not contained in the partial summation, could break the universality, it
has been shown independently by the Bethe ansatz solution 
\cite{Andrei83} and by numerical renormalization group (NRG) (for a recent
review see \cite{Bulla08}) that universality in the above sense indeed holds 
for the Kondo problem. 

Universality is the starting point for the renormalization 
group method \index{renormalization group} whose essence we discuss next.
Let all physical quantities $A_n=h_n(\omega/T^*,T/T^*)$ of a system 
depend on energy $\omega$ and temperature $T$ in a universal way, with  
universal functions $h_n$ and some (yet unknown) characteristic scale $T^*$, 
which depends on the microscopic parameters of the Hamiltonian,
$J_0,\,D_0,\,N(0)$. The fact that the $A_n$ depend on these parameters only
implicitly through $T^*$ implies that different values of this 
parameter set realize the same physical system (defined by its observables 
$A_n$), if only the different sets of parameter values lead to 
the same scale $T^*$. 
In particular, systems with low and with high values of the conduction 
bandwidth or cutoff $D_0$ \index{cutoff}
must be equivalent if the coupling constant $J_0$ is 
adjusted appropriately. In the Kondo problem we are mostly interested 
in the low-energy behavior, where the perturbation theory fails. This 
regime corresponds to electrons with a low bandwidth, 
scattering near the Fermi energy. By virtue of the above argument, 
this low-energy regime is connected with the high bandwidth regime,
where perturbative calculations are possible. 
In the renormalization group method, this relation is established 
iteratively. Starting from an initial high-energy cutoff $D_0$, the 
cutoff is stepwise reduced to low energies, calculating at each step how the 
coupling constant $J$ of the Hamiltonian must be changed,
such that the physical observables $A_n$ remain constant, see 
Fig.~\ref{fig:RG}~(b). This defines a running cutoff $D$ with initial 
value $D_0$ and a ``renormalized'' or ``running'' coupling constant $J(D)$
with initial value $J_0$. The running coupling constant, as part of the 
Hamiltonian, defines a change of the Hamiltonian itself.
More generally, the cutoff reduction may even generate 
new types of interaction operators in the Hamiltonian, implied by the 
requirement that physical observables be invariant. The repeated operations 
on the Hamiltonian, defined in this way, form a semigroup (without 
existence of the inverse operation), the renormalization group (RG). 
The change of the Hamiltonian by the successive cutoff reduction is called
renormalization group flow. 

We can now perform the renormalization of the Kondo Hamiltonian 
(or coupling constant $J$) explicitly in a perturbative way, 
\index{renormalization group!perturbative} following Anderson \cite{Anderson70}.
To that end, it is convenient to introduce the dimensionless, 
bare coupling $g_0=N(0)J_0$ and running coupling $g=N(0)J$. 
We also define the projector $P_{\delta D}$ of the conduction electron 
energy onto the intervals $[-D,-D+\delta D] \cup [D-\delta D,D]$ 
by which the conduction band is reduced in one RG step as well as the projector 
$(1\hspace*{-0.1cm}-\hspace*{-0.1cm}P_{\delta D})$ onto the 
remaining conduction energy interval, c.f. Fig.~\ref{fig:RG}~(b). 
To impose the invariance of physical 
quantities under the RG flow, it is sufficient to keep the total 
conduction electron-pseudofermion vertex $\hat\gamma_{cf}$ invariant,
since all physical quantities are derived from it within the Kondo model. 
$\hat\gamma_{cf}$ is defined by the following $T$-matrix equation,
\begin{eqnarray}
\hat\gamma_{cf} = 
\hat\gamma_{cf}^{(1)}\,+\,\hat\gamma_{cf}^{(1)}\,G\,\hat\gamma_{cf} .
\label{eq:gamma_tot}
\end{eqnarray}
Here, the bare vertex $\hat\gamma_{cf}^{(1)}$ is defined as in \Eq{eq:gamma1},
$G$ denotes schematically the product of 
$G_c$ and $G_f$ propagators connecting two bare vertices $\hat\gamma_{cf}^{(1)}$
in the direct and exchange diagrams (c.f. Fig.~\ref{fig:PT}), 
and integration over the conduction electron energy in $G$ is 
implied. \Eq{eq:gamma_tot} can be rewritten as,
\begin{eqnarray}
\hat\gamma_{cf} &=& \hat\gamma_{cf}^{(1)}
\,+\,\hat\gamma_{cf}^{(1)}\,[P_{\delta D}G]\,\hat\gamma_{cf} 
\,+\,\hat\gamma_{cf}^{(1)}\,[(1\hspace*{-0.1cm}-\hspace*{-0.1cm}P_{\delta D})G]\,\hat\gamma_{cf} \label{eq:gamma_tot_renorm}\\
&=& \hat\gamma_{cf}^{(1)}
\,+\,\hat\gamma_{cf}^{(1)}\,[P_{\delta D}G]\,
\left\{\hat\gamma_{cf}^{(1)} 
\,+\,\hat\gamma_{cf}^{(1)}\,[\left(P_{\delta D}+(1\hspace*{-0.1cm}-\hspace*{-0.1cm}P_{\delta D})\right)G]\,\hat\gamma_{cf}\right\} 
\,+\,\hat\gamma_{cf}^{(1)}\,[(1\hspace*{-0.1cm}-\hspace*{-0.1cm}P_{\delta D})G]\,\hat\gamma_{cf}\nonumber \\
&=& \left.\hat\gamma_{cf}^{(1)}\right. ' 
\,+\,\left.\hat\gamma_{cf}^{(1)}\right. '\,
[(1\hspace*{-0.1cm}-\hspace*{-0.1cm}P_{\delta D})G]\,\hat\gamma_{cf} \quad +\quad {\cal O}(P_{\delta D}^2), \nonumber \\
\mathrm{with}\hfill && \nonumber\\
\left.\hat\gamma_{cf}^{(1)}\right. '  &=&  \hat\gamma_{cf}^{(1)}
\,+\,\hat\gamma_{cf}^{(1)}\,[P_{\delta D}G]\,\hat\gamma_{cf}^{(1)} \ =:\ 
\hat\gamma_{cf}^{(1)}\,+\,\delta\hat\gamma_{cf}^{(1)}.
\label{eq:gamma_bare_renorm}
\end{eqnarray}
In the first line of \Eq{eq:gamma_tot_renorm}, the integral over the 
intermediate conduction electron energy has been split into the 
infinitesimal high-energy part $P_{\delta D}$ and the remaining part   
$(1\hspace*{-0.1cm}-\hspace*{-0.1cm}P_{\delta D})$. In the second line,
the high-energy part of the equation has been iterated once, and in the 
third line, only terms up to linear order in $P_{\delta D}$ have been retained, 
and all terms have been appropriately rearranged.  
As seen from the third line, the total vertex $\hat\gamma_{cf}$ 
obeys again a $T$-matrix equation, however with a reduced conduction 
bandwidth, $(1\hspace*{-0.1cm}-\hspace{-0.1cm}P_{\delta D})$. Moreover, 
by this procedure $\hat\gamma_{cf}$ remains invariant, exactly if 
the bare vertex is changed to $\hat\gamma_{cf}^{(1)}$ as defined in 
\Eq{eq:gamma_bare_renorm}. This is the vertex renormalization we are 
seeking. Note that this expression is perturbative, because in 
\Eq{eq:gamma_tot_renorm} we have iterated the $T$-matrix equation only 
once (1-loop approximation). Higher-order iterations, leading to 
higher-order renormalizations in $\hat\gamma_{cf}^{(1)}$ are possible.   
Note that the vertex renormalization $\delta\hat\gamma_{cf}^{(1)}$ in 
\Eq{eq:gamma_bare_renorm} corresponds just to the 2nd-order 
perturbation theory expression calculated in \Eq{eq:PT_log}, see also 
Fig.~\ref{fig:PT}. Thus, 
one can now read off from these equations the renormalization of the 
dimensionless coupling constant $g$ under cutoff reduction $-\delta D$ as,
\begin{eqnarray}
dg= -\frac{d}{dD} \left[  g^2 \int_{-D}^{D} d\varepsilon\,
\frac{1-2f(\varepsilon)}{\varepsilon}\right] \delta D = -\frac{2 g^2}{D} \,
\delta D.
\end{eqnarray}
Usually one takes the logarithmic derivative which ensures that the 
differential range $-\delta D$ by which the cutoff is reduced is proportional 
to the cutoff itself: $\delta D=Dd(\ln D)$. Thus,
\begin{equation}
\frac{dg}{d\ln D} = - 2 g^2
\label{eq:RGequation0}
\end{equation}
This is the differential renormalization group equation (of 1-loop order).
The function on the right-hand side, $\beta(g)=-2g^2$, 
which controls the running coupling constant renormalization, 
is called the $\beta$-function of the RG. \index{$\beta$ function}
The RG equation can be integrated in a straight-forward way with the 
initial condition $g(D_0)=g_0$ to give,  
\begin{equation}
g(D) = \frac{g_0}{1+2g_0\ln (D/D_0)}.
\end{equation}
It is seen that this solution becomes again divergent for 
antiferromagnetic $g_0>0$ when the running 
cutoff reaches the Kondo scale, $D\to T_K=D_0\exp\left[-1/(2g_0)\right]$,
a consequence of the perturbative RG treatment above. However, this 
divergence allows the conclusion that the ground state of the 
single-impurity Kondo model is a spin-singlet state between the impurity spin
and the spin cloud of the surrounding conduction electron spins as 
outlined in the introduction. Moreover, it allows for a more general 
definition of the Kondo spin screening scale $T_K$, namely the value
of the running cutoff $D$ where the coupling constant diverges and the 
singlet starts to be formed. This will be used for the analysis of the 
Kondo-RKKY interplay in section~\ref{sec:Kondo-RKKY}.


\section{RKKY interaction in paramagnetic and half-metals}
\label{sec:RKKY}

\begin{figure}[t]
\centering
\includegraphics[width=0.9\linewidth]{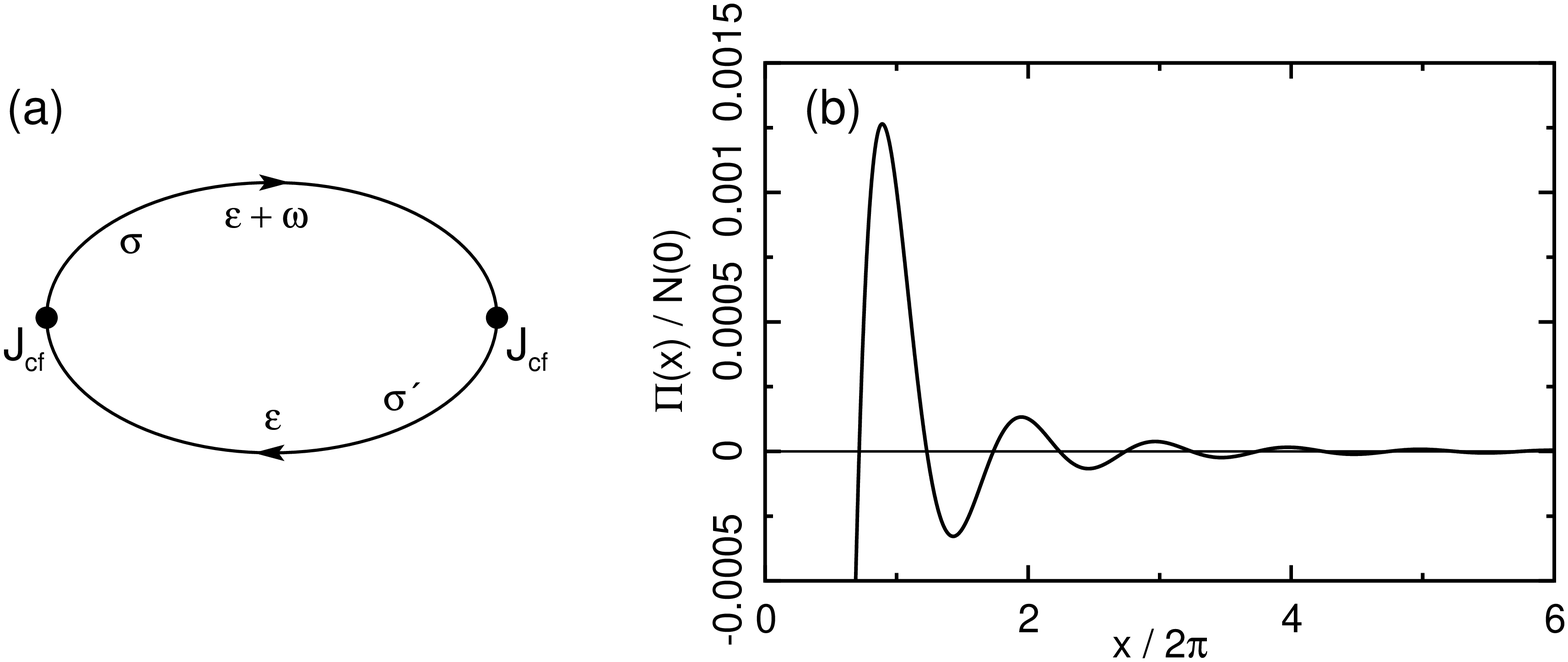}
\caption{
(a) Diagram for the spin-dependent conduction electron polarization function 
$\Pi_{ij}^{\sigma\sigma'}(\omega)$, generating the RKKY interaction.
The solid lines represent conduction electron propagators.
(b) Oscillatory behavior of $\Pi_{ij}^{\sigma\sigma'}(0)$ in a 
paramagnetic metal with isotropic dispersion as a function of 
distance $x=2k_F|{\bf r}_i-{\bf r}_j|$, \Eq{eq:chic_ret} 
\label{fig:diagram_RKKY}
}
\end{figure}

In this section we derive the expressions for the RKKY interaction.
\index{RKKY interaction}
To be general, we will allow for an arbitrary spin polarization of the 
conduction band and then specialize for the paramagnetic case 
(vanishing magnetization) and the half-metallic case (complete
magnetization).  
Thus, we consider now the Kondo lattice Hamiltonian of localized spins 
$\hat{\mathbf{S}_i}$ at the lattice positions ${\bf r}_i$.
\begin{equation}
H = \sum_{{\bf k},\sigma} 
\varepsilon_{\bf k} \; c_{{\bf k} \sigma}^\dagger c_{{\bf k} \sigma}^\phdagger +
J_0 \sum_{i} \hat{{\bf S}}_i \cdot \hat{{\bf s}}_i 
\label{eq:KondoHamiltonian_lattice}
\end{equation}
Usually, the static limit is considered in order to derive a Hamiltonian 
coupling operator. We will later consider the question of 
dynamical correlations as well, as it arises in the interplay with 
the Kondo effect. The interaction Hamiltonian for the conduction 
electrons and the localized $f$-spin $\ve{S}_j$ at a site $j\neq i$, 
\begin{eqnarray}\label{eqn:Hcf}
H^{(cf)}_j=  J_{0}\, \hat{\ve{S}}_{j}\cdot \hat{\ve{s}}_{j} \ ,
\end{eqnarray}
acts as a perturbation for the localized $f$-spin at a site $i$ 
(and vice versa). 
Performing standard thermal perturbation theory by expanding the 
time evolution operator in the interaction picture, 
$\hat T \exp [-\int_0^\beta d\tau H^{cf}_{j}(\tau)]$ up to linear order in $J_0$,
one obtains for the interaction operator of the $f$-spin at site $i$,
up to ${\cal O}(J_0^2)$,  
\begin{eqnarray}\label{eq:HRKKY1}
H^{(2)}_{ij}= J_{0}\, \hat{\ve{S}}_{i}\cdot \hat{\ve{s}}_{i} 
\,-\, J_{0}^2 \left.\langle 
(\hat{\ve{S}}_{i}\cdot \hat{\ve{s}}_{i}) (\hat{\ve{S}}_{j}\cdot \hat{\ve{s}}_{j})
\rangle_{_c}\right|_{\omega=0} \ .
\end{eqnarray}
Here,  $\langle (\dots ) \rangle_{_c} := 
{\rm tr}_c\{{\rm e}^{-\beta H} (\dots ) \}/Z_{G}$, denotes the thermal trace over 
the conduction electron Hilbert space, and the static limit, $\omega=0$, 
has been taken. 
Using Wick's theorem with respect to the conduction electron operators, 
the second term in \Eq{eq:HRKKY1} can be written as,
\begin{eqnarray}\label{eq:HRKKY2}
H^{RKKY}_{ij} \hspace*{-0.2cm}&=&\hspace*{-0.2cm} - \frac{J_{0}^2}{4} \sum_{\alpha,\beta = x,y,z}\sum_{\sigma\sigma'}  
\hat{S}_{i}^{\alpha}\, 
\sigma^{\alpha}_{\sigma\sigma'}\sigma^{\beta}_{\sigma'\sigma}\, 
\hat{S}_{j}^{\beta}  \Pi_{ij}^{\sigma\sigma'}(0),\nonumber\\ &&
\end{eqnarray}
where $\hat{S}^{\alpha}_{i}$, $\alpha=x,\,y,\,z$, are the components of the 
impurity spin, $\sigma^{\alpha}$ the Pauli matrices, and 
$\Pi_{ij}^{\sigma\sigma'}$ the conduction electron density propagator 
between the sites $i$ and $j$ as depicted diagrammatically 
in Fig.~\ref{fig:diagram_RKKY} (a). It has the general form,
\begin{eqnarray} 
\Pi_{ij}^{\sigma\sigma'}(i\omega) = -\frac{1}{\beta}\sum_{\varepsilon_n}
G_{ji\,\sigma}(i\varepsilon_n+i\omega) G_{ij\,\sigma'}(i\varepsilon_n) \ .
\label{eq:Pi_general} 
\end{eqnarray}
In the static limit it reads, 
\begin{eqnarray}
\Pi_{ij}^{\sigma\sigma'}(0) = - \int d\varepsilon\  f(\varepsilon)\,
\hspace*{-0.1cm}\left[
A_{ij\,\sigma}(\varepsilon) {\rm Re} G_{ij\,\sigma'}(\varepsilon) +
A_{ij\,\sigma'}(\varepsilon) {\rm Re} G_{ij\,\sigma}(\varepsilon) 
\right] \,,  \nonumber
\end{eqnarray}
where $A_{ij\,\sigma}(\varepsilon) = -{\rm Im}G_{ij\,\sigma}(\varepsilon +i0)/\pi$. 
Performing the spin contractions in Eq.~(\ref{eq:HRKKY2}) and 
defining the longitudinal and the transverse polarization functions,
respectively, as
\begin{eqnarray}
\Pi_{ij}^{||}(0)   &=& \frac{1}{2}\sum_{\sigma} \Pi_{ij}^{\sigma\sigma}(0)
= - \sum_{\sigma}\int d\varepsilon\,  f(\varepsilon)\,
A_{ij\,\sigma}(\varepsilon) {\rm Re} G_{ij\,\sigma}(\varepsilon)\\
\Pi_{ij}^{\perp}(0) &=& \frac{1}{2}\sum_{\sigma} \Pi_{ij}^{\sigma\,-\sigma}(0) 
= - \sum_{\sigma}\int d\varepsilon\,  f(\varepsilon)\,
A_{ij\,\sigma}(\varepsilon) {\rm Re} G_{ij\,-\sigma}(\varepsilon) \ , 
\end{eqnarray}
one obtains the RKKY interaction Hamiltonian,  
\begin{eqnarray}\label{eqn:HRKKY3}
H^{RKKY} = \sum_{i\neq j} H^{RKKY}_{ij} 
= - \sum_{i,j} \left[ K_{ij}^{||}\,  \hat{S}_i^z \hat{S}_j^z 
             + K_{ij}^{\perp}\, \left(\hat{S}_i^x\hat{S}_j^x + 
                                   \hat{S}_i^y\hat{S}_j^y\right)\right] 
\nonumber
\end{eqnarray}
\vspace*{0.2cm}\noindent
where the sums run over all (arbitrarily distant) 
lattice sites $i,\,j, i\neq j$, of localized spins 
$\hat{\ve{S}}_i$ and $\hat{\ve{S}}_j$, and
\begin{eqnarray}\label{eqn:KRKKY}
K_{ij}^{||}   =   \frac{1}{2} J_{0}^2  \Pi_{ij}^{||}(0) \ , \qquad
K_{ij}^{\perp} = \frac{1}{2} J_{0}^2 \Pi_{ij}^{\perp}(0) \ , 
\end{eqnarray}
are the longitudinal and transverse RKKY couplings,
respectively. 
The spin being a vector operator, the interaction Hamiltonian 
$H^{RKKY}_{ij}$ has a tensor structure and is, in general, anisotropic
for a magnetized conduction band, as seen from Eq.~(\ref{eqn:KRKKY}).

We now present explicitly the expressions for the special cases of a 
paramagnet and of a half-metal. For a paramagnetic conduction band we have
$G_{ij\,\sigma}=G_{ij, -\sigma}$, independent of spin. Hence, the RKKY coupling is
isotropic, and we have the paramagnetic RKKY Hamiltonian, 
\begin{eqnarray}\label{eqn:HRKKY_PM}
H^{RKKY}_{PM} = - \sum_{(i,j)} K_{ij}^{PM}\, \hat{\ve{S}}_i\cdot\hat{\ve{S}}_j  \ ,
\end{eqnarray}
with 
\begin{eqnarray}
K_{ij}^{PM} = -\frac{J_{0}^2}{2}\sum_{\sigma} 
\int d\varepsilon\, f(\varepsilon)\, 
A_{ij\,\sigma}(\varepsilon) {\rm Re} G_{ij\,\sigma}(\varepsilon) \ .
\phantom{xxx}
\label{eqn:JRKKY_PM}
\end{eqnarray}
For a half-metal, i.e., for a completely spin-magnetized conduction band
with majority spin $\sigma=\uparrow$ we have $A_{ij\,\downarrow}(\varepsilon)=0$, 
Thus, the half-metallic RKKY Hamiltonian reads,
\begin{eqnarray}\label{eqn:HRKKY_FM}
H^{RKKY}_{FM} = - \sum_{(i,j)} \, \left[
K_{ij}^{FM\ ||}\, \hat{S}_i^z\hat{S}_j^z  
 + K_{ij}^{FM\ \perp}\, \left(\hat{S}_i^x\hat{S}_j^x+\hat{S}_i^y\hat{S}_j^y
\right)\right] \ , 
\end{eqnarray}
with 
\begin{eqnarray}
K_{ij}^{FM\ ||} &=& - \frac{J_{0}^2}{2} 
\int d\varepsilon\, f(\varepsilon)\, 
A_{ij\,\uparrow}(\varepsilon) {\rm Re} G_{ij\,\uparrow}(\varepsilon) 
\label{eqn:JRKKY_FM1} \\ 
K_{ij}^{FM\ \perp} &=& - \frac{J_{0}^2}{2} 
\int d\varepsilon\, f(\varepsilon)\, 
A_{ij\,\uparrow}(\varepsilon) {\rm Re} G_{ij\,\downarrow}(\varepsilon) \ . 
\phantom{xxxxx}
\label{eqn:JRKKY_FM2} 
\end{eqnarray}
The missing spin summation in Eqs.~(\ref{eqn:JRKKY_FM1}), (\ref{eqn:JRKKY_FM2}) 
as compared to Eq.~(\ref{eqn:JRKKY_PM}) indicates that 
in the completely magnetized band only the majority spin species contributes
to the coupling.
Note, however, that the transverse coupling $K^{FM\perp}_{ij}$ is still 
non-zero even in the ferromagnetically saturated case because of virtual
(off-shell) minority spin contributions represented by the real part 
${\rm Re} G_{ij\,\downarrow}(\varepsilon)$ in Eq.~(\ref{eqn:JRKKY_FM2}).

The RKKY coupling is long-ranged and has in general complex,
oscillatory behavior in space, because it depends on details of the 
conduction band structure via the position dependent 
Green's functions $G_{ji\,\sigma}(\omega)$ in \Eq{eq:Pi_general}.
For an isotropic system in $d=3$ dimensions, the retarded/advanced
conduction electron Green's function $G_{r\sigma}(\varepsilon\pm i0)$ 
and the paramagnetic polarization $\Pi_{r}^{\sigma\sigma'}(\omega)$ 
at temperature $T=0$ are calculated in position space as,
\begin{eqnarray}
G_{ r\sigma}(\varepsilon \pm i0) &=& - \pi N(\varepsilon)\,
\frac{{\rm e}^{\pm i k(\varepsilon_F+\varepsilon)r}}
{k(\varepsilon_F+\varepsilon)r} \\
\Pi_{r}^{\sigma\sigma'}(\omega+i0) &=& 
\ \ \left[  N(0)\,
\frac{\sin(x)-x\cos(x)}{4x^4}\ + 
\ {\cal O}\left(\left(\frac{\omega}{\varepsilon_F}\right)^2\right)
\right] \label{eq:chic_ret} \\
&\pm& i \left[\ \frac{1}{\pi} N(0)\, \frac{1-\cos(x)}{x^2}\, 
\frac{\omega}{\varepsilon_F}\ 
+ \ {\cal O}\left(\left(\frac{\omega}{\varepsilon_F}\right)^3\right)
\ \right]  \nonumber 
\end{eqnarray} 
Here, $\varepsilon_F$ and $k_F$ are the Fermi energy and Fermi wavenumber, 
respectively, and $r=|{\bf r}_i-{\bf r}_j|$, $x=2k_Fr$. 
For illustration, Fig.~\ref{fig:diagram_RKKY} (b) shows the static polarization
$\Pi_{r}^{\sigma\sigma'}(0)$ as a function of $x$ for the isotropic case.


\section{Interplay of Kondo screening and RKKY interaction}
\label{sec:Kondo-RKKY}

We now turn to the interplay of the two interactions on a Kondo lattice, 
\Eq{eq:KondoHamiltonian_lattice}. 
First, it is crucial to remember that the RKKY interaction between 
different $f-$spins is not a direct spin
exchange interaction, but mediated by the conduction band 
\cite{Ruderman54,Kasuya56,Yosida57} and generated in second 
order by the same spin coupling $J_0$ that is also responsible for 
the local Kondo spin screening, as shown in the previous 
section \ref{sec:RKKY}. 
The essential difference can be seen from the example of a 
two-impurity Kondo system, ${\bf S}_1$, ${\bf S}_2$: 
\index{two-impurity Kondo problem}
With a direct impurity-impurity coupling, $K\, {\bf S}_1 \cdot {\bf S}_2 $, 
this model can exhibit a dimer singlet phase where the 
dimer is decoupled from the conduction electrons. The dimer singlet and the 
local Kondo singlet phase are then separated by a quantum critical 
point (QCP), controlled by $K$ \cite{Jones88,Affleck95}.   
By contrast, when the interimpurity coupling is created by the RKKY 
interaction only, i.e. generated by $J_0$, a decoupled dimer singlet phase 
is not possible. Instead, the impurity spins must remain coupled to the 
conduction sea. We show below that the Kondo singlet formation at $T=0$ 
breaks down at a critical strength of the RKKY coupling even if magnetic 
ordering is suppressed, i.e. without a 2nd-order quantum phase transition 
and without critical fluctuations. If magnetic ordering occurs, 
critical ordering fluctuations will be present in addition to, 
but independently of the RKKY-induced Kondo breakdown.

\subsection{The concept of a selfconsistent renormalization group}
\label{subsec:Kondo-RKKY_RG}

The problem of local Kondo screening or breakdown \index{Kondo breakdown}
on a Kondo lattice amounts to calculating the 
vertex for scattering of $c$-electrons from a local $f$-spin and analyzing 
its divergence (Kondo screening of the $f$-spin, c.f. 
section~\ref{subsec:RG}) or non-divergence (Kondo breakdown) under RG.
In the case of multiple Kondo sites or a Kondo lattice, this vertex
$\hat\Gamma_{cf}$ acquires nonlocal contributions in addition to the 
local coupling $J_0$, because a $c-$electron can scatter 
from a distant Kondo site $j\neq i$, and the spin flip at that site is 
transferred to the $f-$spin at site $i$ via the RKKY interaction.
On the other hand, the RKKY vertex $\hat\Gamma_{ff}$ coupling two $f-$spins 
has no logarithmic RG flow, since the recoil (momentum integration) 
of the itinerant conduction electrons prevents an infrared divergence 
of the RKKY interaction. Thus, $\hat\Gamma_{ff}$ remains in the weak 
coupling regime, and RKKY-induced magnetic ordering must be a secondary 
effect, not controlled by the RG divergence of a coupling constant.

\begin{figure}[t]
\centering
\includegraphics[width=0.80\linewidth]{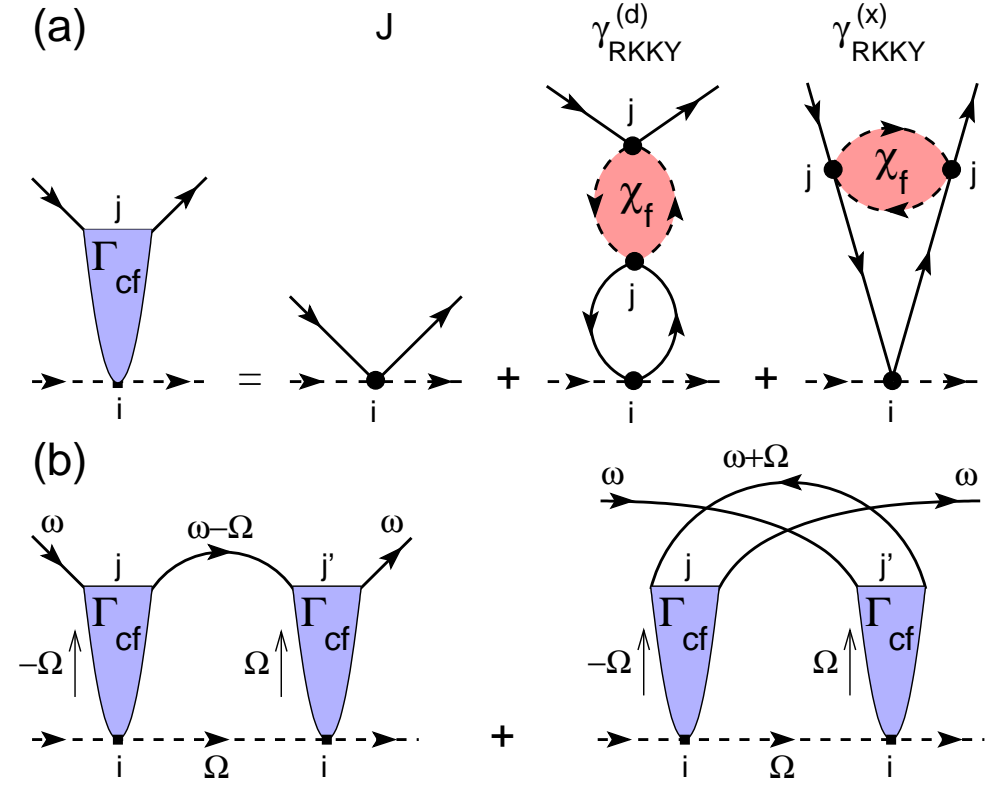}
\caption{\label{fig:vertex} 
(a) $f$-spin--$c$-electron vertex $\hat\Gamma_{cf}$, composed of the onsite 
vertex $J$ at site $i$ and the RKKY-induced contributions from 
surrounding sites $j\neq i$ to leading order in the RKKY coupling, 
$\gamma_{RKKY}^{(d)}$ (direct term) and $\gamma_{RKKY}^{(x)}$ (exchange term).   
(b) 1-loop diagrams for the perturbative RG. 
Solid lines: electron Green's functions $G_c$, 
dashed lines: pseudofermion propagators $G_f$ of the local $f-$spins. 
The red bubbles represent the full $f-$spin susceptibility 
at sites $j$.  }
\end{figure}

The diagrams contributing to $\hat\Gamma_{cf}$ to leading order in the 
RKKY coupling are shown in Fig.~\ref{fig:vertex}~(a). 
As seen from the figure, a nonlocal scattering process necessarily 
involves the exact, local dynamical $f$-spin susceptibility $\chi_f(i\Omega_n)$ 
on site $j$. The resulting $c-f$ vertex $\hat\Gamma_{cf}$ has the 
structure of a nonlocal Heisenberg coupling in spin space, see 
Appendix \ref{subsec:RKKYvertex-spin}. 
The exchange diagram, $\gamma_{RKKY}^{(x)}$ in Fig.~\ref{fig:vertex} (a), 
contributes only a subleading logarithmic term as compared to the direct
term $\gamma_{RKKY}^{(d)}$, see Appendix \ref{subsec:RKKYvertex-energy}. 
In particular, it does not alter the universal $T_K(y)$ suppression 
derived below. It can, therefore, be
neglected. To leading (linear) order in the RKKY coupling, 
$\hat\Gamma_{cf}$ thus reads (in Matsubara representation), 
\begin{eqnarray}
\hat\Gamma_{cf} &=& 
\left[ J  \delta_{i,j} + \gamma^{(d)}_{RKKY}({\bf r}_{ij},i\Omega_n)\right] 
{\bf S}_i \cdot {\bf s}_j 
\label{eq:Gammacf}\\ 
&=& \left[ J\delta_{ij}+ 2J J_0^2\ (1-\delta_{ij})\ \Pi({\bf r}_{ij}, i\Omega_n)\, 
\tilde\chi_f(i\Omega_n) \right] 
{\bf S}_i \cdot {\bf s}_j \, ,
\nonumber
\end{eqnarray}
where ${\bf r}_{ij} = {\bf x}_i - {\bf x}_j$ the distance vector 
between the sites $i$ and $j$, and $\Omega$ is the energy transferred in the 
scattering process. $\Pi({\bf r}_{ij}, i\Omega_n)$ is the $c-$electron 
density correlation function between sites $i$ and $j$ [bubble of solid lines 
in Fig.~\ref{fig:vertex} (a)] and 
$\tilde\chi_f(i\Omega_n):=\chi_f(i\Omega_n)/(g_L\mu_B)^2$, with $g_L$ the 
Land\'e factor and $\mu_B$ the Bohr magneton. 
Note that \Eq{eq:Gammacf} contains the running coupling 
$J$ at site $i$ which will be renormalized under RG, 
while at the site $j$, where the $c-$electron scatters, the bare 
coupling $J_0$ appears, since all vertex renormalizations on that 
site are already included in the exact susceptibility $\chi_f$. 
Higher order terms, as for instance generated by the RG
[see below, Fig.~\ref{fig:vertex} (b)], lead to nonlocality of the incoming 
and outgoing coordinates of the scattering $c-$electrons, 
${\bf x}_{j}$, ${\bf x}_{j'}$, but the $f$-spin coordinate ${\bf x}_i$ 
remains strictly local, since the pseudofermion propagator 
$G_f(i\nu_n)=1/i\nu_n$ is local \cite{Kroha98}. 
For this reason, speaking of Kondo singlet formation on 
a single Kondo site is well defined even on a Kondo lattice, and so is 
the local susceptibility $\chi_f$ of a single $f$-spin.
The corresponding Kondo scale $T_K$ on a site $j$ is observable, 
e.g., as the Kondo resonance width measured by STM spectroscopy on one Kondo 
ion of the Kondo lattice.
The temperature dependence of the single-site $f$-spin susceptibility 
is known from the Bethe ansatz solution \cite{Andrei83} in terms of the 
Kondo scale $T_K$. \index{Bethe ansatz solution} It has a 
$T=0$ value $\chi_f(0)\propto 1/T_K$ and crosses over to the $1/T$ 
behavior of a free spin for $T>T_K$. These features can be 
modeled in the retarded/advanced, local, dynamical  
$f-$spin susceptibility $\chi_f(\Omega\pm i0)$ as,
\begin{eqnarray}
\chi_f(\Omega \pm i0) &=& \frac{(g_L\mu_B)^2W}{\pi T_K\,\sqrt{1+(\Omega/T_K)^2}}\, 
\left( 1 \pm \frac{2i}{\pi} {\rm arsinh} \frac{\Omega}{T_K} \right)
\label{eq:chif_ret}
\end{eqnarray} 
where $W$ is the Wilson ratio, and the imaginary part is implied by the 
Kramers-Kronig relation. 

Deriving the one-loop RG equation for a multi-impurity or lattice 
Kondo system proceeds as in section~\ref{subsec:RG}, however for 
the $c-f$ vertex $\hat\Gamma_{cf}$, including RKKY-induced, nonlocal 
contributions. The one-loop spin vertex function is shown diagrammatically 
in Fig.~\ref{fig:vertex} (b). Using Eq.~(\ref{eq:Gammacf}), the sum of these
two diagrams is up to linear order in the RKKY coupling, 
\begin{eqnarray}
Y({\bf r}_{ij}, i\omega_n ) &=&   \label{eq:Yx}   
-J \, T \sum_{i\Omega_m}   
\left[  J\delta_{ij} +
\gamma^{(d)}_{RKKY}({\bf{r}}_{ij},i\Omega_m)+\gamma^{(d)}_{RKKY}({\bf{r}}_{ij},-i\Omega_m)
\right]  \\
&&\hspace*{+1.8cm}
\times 
\left[ 
G_c({\bf r}_{ij}, i\omega_n -i\Omega_m )-G_c({\bf r}_{ij}, i\omega_n + i\Omega_m)
\right]\, G_f(i\Omega_m). 
\nonumber
\end{eqnarray}
Here, $\omega$ is the energy of incoming conduction electrons,    
$G_c({\bf r}_{ij}, i\omega_n + i\Omega_m)$ is the single-particle 
$c-$electron propagator from the incoming to the outgoing site.

For the low-energy physics, the vertex renormalization for $c-$electrons 
at the Fermi surface is required.   
This means setting the energy $i\omega\to\omega =0+i0$  and 
Fourier transforming the total vertex 
$Y({\bf r}_{ij}, i\omega)$ with respect to the incoming and outgoing 
$c-$electron coordinates, ${\bf x}_j$, ${\bf x}_i$, and taking 
its Fourier component for momenta at the Fermi surface, ${\bf k}_F$. 
Note that at the Fermi energy  
$Y({\bf k}_F, 0)$ is real, even though the RKKY-induced, 
dynamical vertex $\gamma^{(d)}_{RKKY}(\pm i\Omega_m)$ appearing in 
\Eq{eq:Yx} is complex-valued. This ensures  
the total vertex operator of the renormalized Hamiltonian to be
Hermitian. By analytic continuation, the Matsubara summation
in \Eq{eq:Yx} becomes an integration over 
the intermediate $c-$electron energy from the lower and upper 
band cutoff $\mp D$ to the Fermi energy ($\Omega=0$). The 
coupling constant renormalization is then obtained  
by requiring that $Y({\bf k}_F, 0)$ is invariant under an infinitesimal 
reduction of the running band cutoff $D$ (c.f. section~\ref{subsec:RG}).
Note that the band cutoff appears in both, the
intermediate electron propagator $G_c$ and in $\Pi$.
However, differentiation of the latter does not contribute to 
the logarithmic RG flow. This leads to the 1-loop lattice RG 
equation for the local coupling \cite{Nejati17}, 
\begin{equation}
\frac{{\rm d}g}{{\rm d}\ln  D} = -2 g^2\ 
\left( 1 - y\, g_0^2 \ \frac{D_0}{T_K}\ \frac{1}{\sqrt{1+(D/T_K)^2}} 
\right) \ ,
\label{eq:RGequation}
\end{equation}
with the bare band cutoff $D_0$. 
The first term in \Eq{eq:RGequation} is the onsite 
contribution to the $\beta-$function, \index{$\beta$ function}
while the second term represents the RKKY contribution. 
It is seen that $\chi_f$, as in \Eq{eq:chif_ret},
induces a soft cutoff \index{cutoff!soft}
on the scale $T_K$ and the characteristic
$1/T_K$ dependence to the RG flow of this contribution, where 
$T_K$ is the Kondo scale on the {\it surrounding} Kondo sites.      
The dimensionless coefficient 
\begin{equation}
y=-\frac{8W}{\pi ^2} {\rm Im} \sum_{j\neq i}
\frac{{\rm e}^{-i{\bf k}_F{\bf r}_{ij}}} {N(0)^2}
 G_c^R({\bf r}_{ij}, \Omega=0 ) \,
\Pi({\bf r}_{ij}, \Omega=0 ) 
\label{eq:RKKYparameter1}
\end{equation}
arises from the Fourier transform $Y({\bf k}_F, 0)$ and  
parameterizes the RKKY coupling strength. The summation 
in \Eq{eq:RKKYparameter1} runs over all positions $j\neq i$ 
of Kondo sites in the system. It is important to note that  
$y$ is generically positive, even though the RKKY correlations 
$\Pi({\bf r}_{ij},0)$ may be ferro- or antiferromagnetic. 
For instance, for an isotropic and dense system with lattice constant 
$a$ ($k_Fa \ll 1$), 
the summation in Eq.~(\ref{eq:RKKYparameter1}) can be approximated 
by an integral, and with the substitution $x=2k_F|{\bf r}_{ij}|$, $y$ 
can be expressed as 
\begin{equation}
y \approx  \frac{2W}{(k_Fa)^3}
\int_{k_Fa}^{\infty} dx\ (1-\cos x)\ \frac{x \cos x - \sin x}{x^4} > 0 \ .
\label{eq:RKKYparameter2}
\end{equation} 
As a consequence, the RKKY correlations reduce the $g-$renormalization 
in \Eq{eq:RGequation}, irrespective of the sign of $\Pi({\bf r}_{ij},0)$,
as one would physically expect. 

The Kondo scale for singlet formation on site $i$ 
is defined as the running cutoff value where the $c-f$ coupling $g$ diverges.
An important feature of the lattice RG equation (\ref{eq:RGequation}) 
is that the Kondo screening scale on surrounding sites $j\neq i$ appears as a 
parameter in the $\beta$-function for the renormalization on site $i$.   
By equivalence of all Kondo sites, the Kondo scales $T_K$ on all sites 
$i$ and $j$ must be equal. This leads to the fact that the divergence 
scale $T_K$ of the lattice RG equation must be determined 
selfconsistently \index{selfconsistency} and will imply  
an implicit equation for the local screening scale $T_K=T_K(y)$ 
on a Kondo lattice, which will depend  
on the RKKY parameter $y$.
The equivalence of the $c-f$ vertices on all Kondo sites is reminiscent 
of a dynamical mean-field theory treatment, 
\index{selfconsistency!and dynamical mean field theory} 
however, it goes beyond the latter in taking the long-range RKKY contributions 
into account.

\subsection{Integration of the RG equation}
\label{subsec:RGintegration}

The RG equation \Eq{eq:RGequation} is readily integrated by separation 
of variables,
\begin{eqnarray}
-\int_{g_0}^{g}\frac{dg}{g^2} \ = \
2 \int_{\ln D_0}^{\ln D} d\ln D' 
\label{eq:RGint1} 
\,-\,2yg_0^2\frac{D_0}{T_K}\int_{D_0/T_K}^{D/T_K}\frac{dx}{x}
\frac{1}{\sqrt{1+x^2}} \ ,
\end{eqnarray}
 
or 
\begin{eqnarray}
\hspace*{0.5cm} \frac{1}{g}-\frac{1}{g_0} \ =\  \label{eq:RGint2}
2 \ln\left(\frac{D}{D_0}\right) 
\, -\,  yg_0^2\frac{D_0}{T_K} 
\ln\left(
\frac{\sqrt{1+(D/T_K)^2}-1}{\sqrt{1+(D/T_K)^2}+1}
\right) \ ,
\end{eqnarray} 
where we have used $D_0/T_K\gg 1$ in the last expression.
The Kondo scale is defined as the value of the running cutoff $D$
where $g$ diverges, i.e., $g\to\infty$  when $D\to T_K$. This 
yields the defining equation for the Kondo scale $T_K\equiv T_K(y)$,
\begin{eqnarray}
-\frac{1}{g_0} \ = \ \label{eq:TKydef1}
2 \ln\left(\frac{T_K(y)}{D_0}\right) 
\,-\,yg_0^2\frac{D_0}{T_K(y)} 
\ln\left(
\frac{\sqrt{2}-1}{\sqrt{2}+1}
\right) .
\nonumber
\end{eqnarray} 
Using the definition of the single-impurity Kondo temperature,
$-1/g_0=2\ln\left(T_K(0)/D_0\right)$, the defining, implicit equation 
for $T_K(y)$ can finally be written as 
\begin{equation}
\frac{T_K(y)}{T_K(0)} = 
\exp \left( -y\, \alpha\, g_0^2\, \frac{D_0}{T_K(y)} \right) \ ,
\label{eq:TK_y}
\end{equation}
with \hbox{$\alpha =  \ln (\sqrt{2}+1)$}.

\subsection{Universal suppression of the Kondo scale} 
\label{subsec:universality}

\begin{figure}[t] 
\centering
\includegraphics[width=1.0\linewidth]{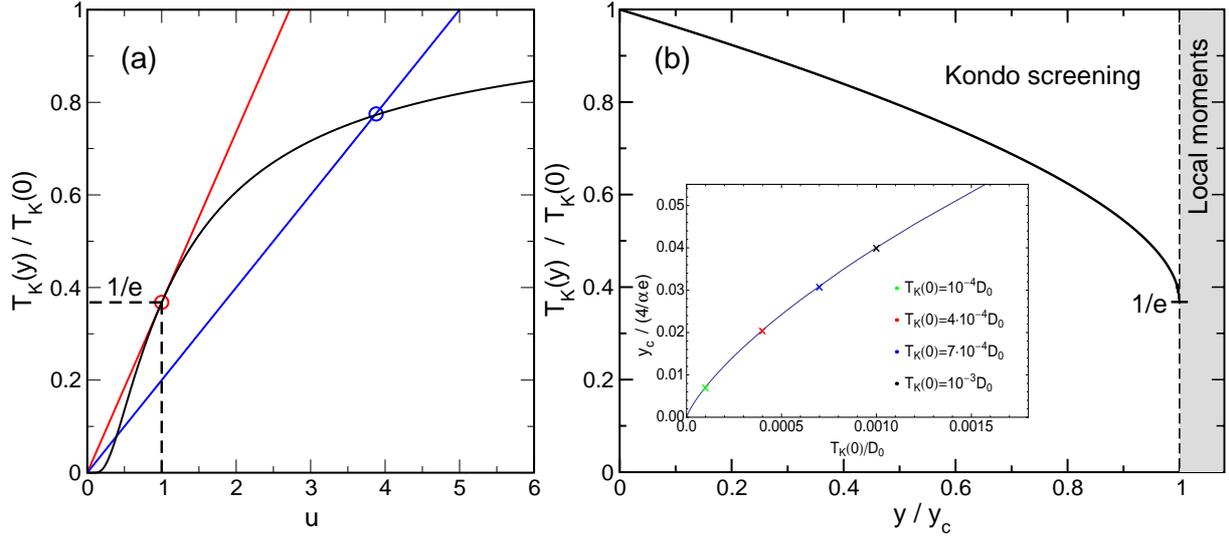}
\caption{\label{fig:TK_y_universal}
(a) Graphical solution of \Eq{eq:TK_y_universal}: 
black, solid curve: right-hand side of \Eq{eq:TK_y_universal}, blue line: 
left-hand side for $y<y_c$, red line: left-hand side for $y=y_c$ (where 
red line and black curve touch). It proofs that there is a
critical coupling $y_c$ beyond which \Eq{eq:TK_y_universal} has no solution,
and $T_K(y_c)/T_K(0)=1/e$.
(b) Universal dependence of $T_K(y)/T_K(0)$ on the normalized 
RKKY parameter $y/y_c$, solution of \Eq{eq:TK_y_universal}. 
The inset shows the critical RKKY parameter $y_c$ for various single-ion
Kondo temperatures $T_K(0)$, \Eq{eq:yc}.
}   
\end{figure}

By the rescaling,
$u=T_K(y)/(y\alpha g_0^2D_0)$, $y_c=T_K(0)/(\alpha {\rm e} g_0^2D_0)$, 
\Eq{eq:TK_y} takes the universal form ({\rm e} is Euler's constant),
\begin{equation}
\frac{y}{{\rm e}y_c}\, u={\rm e}^{-1/u} \ .
\label{eq:TK_y_universal}
\end{equation}
Its solution can be expressed in terms of the Lambert $W$ function 
\cite{Lambert} as $u(y)=-1/W(-y/{\rm e}y_c)$. 
Fig.~\ref{fig:TK_y_universal}~(a)
visualizes solving \Eq{eq:TK_y_universal} graphically. It shows 
that \Eq{eq:TK_y_universal} has solutions only for $y\leq y_c$. This means that 
$y_c$ marks a Kondo breakdown point beyond which the RG does not scale 
to strong coupling, i.e., a Kondo singlet is not formed for $y>y_c$ 
even at the lowest energies. 
Using the above definitions, the RKKY-induced suppression of the Kondo 
lattice temperature reads, 
$T_K(y)/T_K(0)= u(y)y/({\rm e}y_c)=-y/[{\rm e}y_cW(-y/{\rm e}y_c)]$. It is
shown in Fig.~\ref{fig:TK_y_universal}~(b). In particular, at the breakdown 
point it vanishes {\it discontinuously} and takes the finite, 
universal value (see Fig.~\ref{fig:TK_y_universal}~(a)),
\begin{eqnarray}
\frac{T_{K}(y_c)}{T_K(0)}=\frac{1}{{\rm e}}\approx 0.368\ . 
\label{eq:TK_yc}
\end{eqnarray}
We emphasize that the RKKY parameter $y$ depends on details of the conduction 
band structure and of the spatial arrangement of Kondo sites.
Subleading contributions to $\Gamma_{cf}$ may modify the form of the cutoff 
function in the RG Eq.~(\ref{eq:RGequation}) and thus the nonuniversal 
parameter $\alpha$. However, all this does not affect the universal 
dependence $T_K(y)$ on $y$ given by \Eq{eq:TK_y_universal}.

The critical RKKY parameter, as defined before \Eq{eq:TK_y_universal}, 
can be expressed solely in terms of the single-ion Kondo scale, 
\begin{equation}
y_{c} = \frac{4} {\alpha{\rm e}}\tau_K ({\rm ln}\tau_K)^2\ ,
\label{eq:yc}
\end{equation}
with $\tau_K=T_K(0)/D_0$. 
Note that [via $T_K(0)=D_0 \exp (-1/2g_0)$ and $N(0)=1/(2D_0)$] this is 
equivalent to Doniach's breakdown criterion \cite{Doniach77}, 
$N(0)y_cJ_0^2 = T_K(0)$, up to a factor of ${\cal O}(1)$. However, the present 
theory goes beyond the Doniach scenario in that it predicts the 
behavior of $T_K(y)$. 
 
\begin{figure}[t]
\centering
\includegraphics[width=0.6\linewidth]{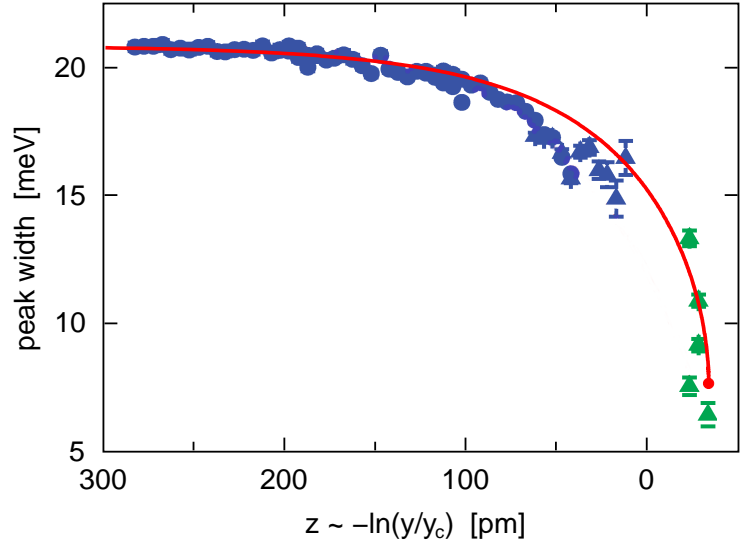}
\caption{\label{fig:TK_y_exp_th} Comparison of the theory (red curve), 
\Eq{eq:TK_y_universal}, with STM spectroscopy experiments on a tunable 
two-impurity Kondo system (data points, Ref.~\cite{Bork11}). 
The data points represent the Kondo scale $T_K$ as extracted from the STM
spectra by fitting a split Fano line shape of width $T_K$ to the 
experimental spectra, see \cite{Bork11} for experimental details.} 
\end{figure}

The present theory applies directly to two-impurity 
Kondo systems, where magnetic ordering does not play a role, 
and can be compared to corresponding STM experiments 
\cite{Bork11,Kroeger11}. \index{STM spectroscopy}
In Ref. \cite{Bork11}, the Kondo scale has been extracted as the 
line width of the (hybridization-split) Kondo-Fano resonance. In this 
experimental setup, the RKKY parameter $y$ is proportional to the 
overlap of tip and surface $c-$electron wave functions and, thus,
depends exponentially on the tip-surface separation $z$, 
$y=y_c\,\exp(-(z-z_0)/\xi)$. Identifying the experimentally observed
breakdown point, $z=z_0$, with the Kondo breakdown point, 
the only adjustable parameters are a scale factor $\xi$ of the $z$ coordinate
and $T_K(0)$, the resonance width at large separation, $z=300$~pm. 
The agreement between theory and experiment is striking, as shown in
Fig.~\ref{fig:TK_y_exp_th}. In particular, at the breakdown point
$T_K(y_c)/T_K(0)$ coincides accurately with the prediction, 
\Eq{eq:TK_yc}, without any adjustable parameters.


\section{Conclusion}
\label{sec:conclusion}

We have derived a perturbative renormalization group theory 
for the interference of Kondo singlet formation and 
RKKY interaction in Kondo lattice and multi-impurity systems,
assuming that magnetic ordering is suppressed, e.g. by frustation. 
Eqs.~(\ref{eq:TK_y}) or (\ref{eq:TK_y_universal}) represent a mathematical 
definition of the energy scale for Kondo singlet formation in a Kondo lattice,
i.e., of the Kondo lattice temperature, $T_K(y)$.
The theory predicts a universal suppression of $T_K(y)$ and a breakdown of 
complete Kondo screening at a critical RKKY parameter, $y=y_c$. 
At the breakdown point, the Kondo scale takes a {\it finite}, universal value,
$T_K(y_c)/T_K(0)=1/{\rm e}\approx 0.368$, and vanishes 
{\it discontinuously} for $y>y_c$. 
In the Anderson lattice, by contrast to the Kondo lattice, the locality 
of the $f-$spin does no longer strictly hold, but our approach should 
still be valid in this case. The parameter-free, quantitative agreement 
of this behavior with different spectroscopic experiments 
\cite{Bork11,Kroeger11} strongly supports that the present theory 
captures the essential physics of the Kondo-RKKY interplay. 

The results may have profound relevance for heavy-fermion magnetic QPTs. 
In an unfrustrated lattice, the partially screened 
local moments existing for $y>y_c$ must undergo a second-order 
magnetic ordering transition at sufficiently low temperature. 
This means that the bare $c-$electron correlation or polarization function 
$\Pi$ must be replaced by the full $c$-correlation function $\chi_c$ 
and will imply a powerlaw divergence of the latter in \Eq{eq:Gammacf}. 
We have checked the effect of such a  
magnetic instability, induced either by the ordering of remanent 
local moments or by a $c-$electron SDW instability: 
The breakdown ratio $T_K(y_c)/T_K(0)$ will be altered, but must 
remain nonzero. The reason is that the inflection point of the exponential 
on the right-hand side of \Eq{eq:TK_y_universal} 
(see Fig.~\ref{fig:TK_y_universal}) is not removed by such 
a divergence and, therefore, the solution ceases to exist at a finite 
value of $T_K(y_c)$.
This points to an important conjecture about a possible, new quantum 
critical scenario with Kondo destruction: The Kondo spectral weight may
vanish continuously at the QCP, while the Kondo energy scale $T_K(y)$ 
(resonance width) remains finite.
Such a scenario may reconcile apparently contradictory experimental results
in that it may fulfill dynamical scaling, even  
though $T_K(y_c)$ is finite at the QCP.

\newpage 
\section*{Appendices}
\appendix
\section{f-spin -- conduction electron vertex $\hat\Gamma_{cf}$}
\label{sec:cfvertex}

Here we present some details on the calculation of the 
elementary $c$-electron--$f$-spin vertex $\hat\Gamma_{cf}$ 
It is defined via the Kondo lattice Hamiltonian, 
\begin{equation}
H = \sum_{{\bf k},\sigma} 
\varepsilon_{\bf k} \; c_{{\bf k} \sigma}^\dagger c_{{\bf k} \sigma}^\phdagger +
J_0 \sum_{i} \hat{{\bf S}}_i \cdot \hat{{\bf s}}_i  \ ,
\label{S_Hamiltonian}
\end{equation}
The direct ($d$) and exchange ($x$) parts of the RKKY-induced vertex 
can be written as the product of a distance and energy dependent 
function $\Lambda_{RKKY}^{(d/x)}$ and an operator in spin space, 
$\hat\Gamma^{(d/x)}$,
\begin{equation} 
\hat\gamma_{RKKY}^{(d/x)}= \Lambda_{RKKY}^{(d/x)}({\bf r}_{ij},i\Omega) \,
\hat\Gamma^{(d/x)}
\label{eq:RKKYvertex}
\end{equation}

\subsection{Spin structure}
\label{subsec:RKKYvertex-spin}

\begin{figure}[t]
\centering
(a)\ \ \includegraphics[height=0.27\linewidth]{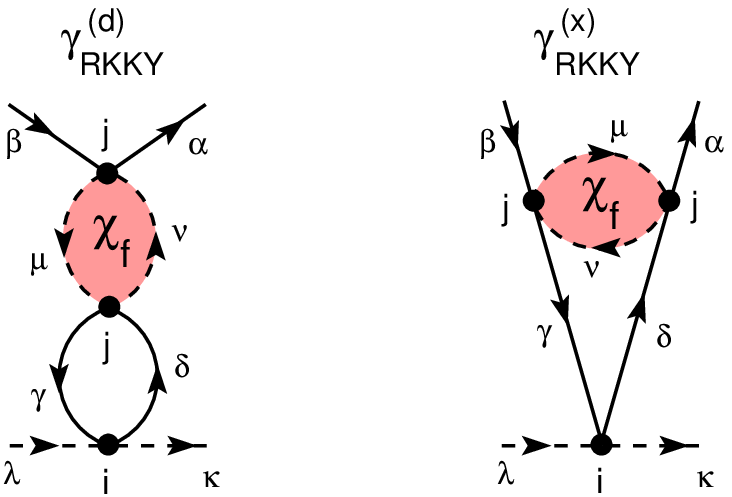}\hfill
(b)\ \ \includegraphics[height=0.27\linewidth]{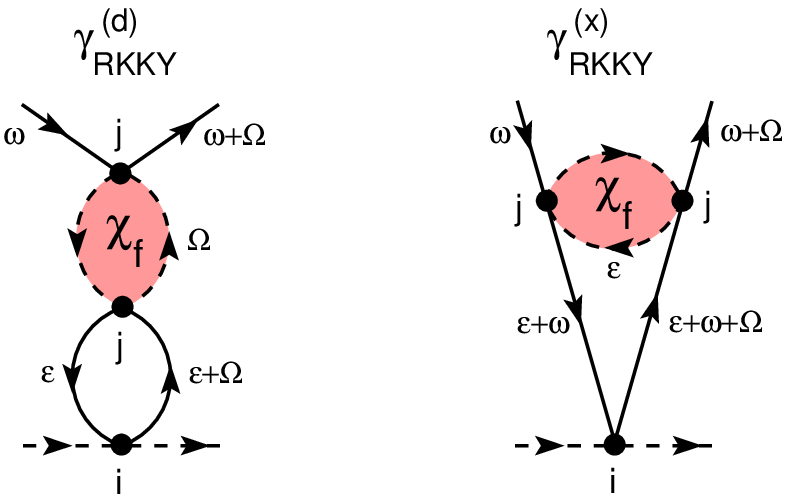}
\caption{\label{fig:RKKYvertex} 
Direct ($d$) and exchange ($x$) diagrams 
of the RKKY-induced contributions to the $c-f$ vertex: (a) 
spin labelling, (b) energy labelling.}
\end{figure}

Denoting the vector of Pauli matrices acting in $c-$electron spin space 
by $\bsig =(\sigma^x,\sigma^y,\sigma^z)^T$ and the vector of Pauli matrices 
in $f-$spin space by ${\bf s}=(s^x,s^y,s^z)^T$,  
the RRKY-induced vertex contributions read in spin space,
\begin{eqnarray}
\hat\Gamma^{(d)}_{\alpha\beta, \kappa\lambda} &=& \hspace*{-0.2cm}
\sum_{a,b,c=x,y,z}\ \sum_{\gamma,\delta,\mu,\nu=1}^2
\left( \sigma^{a}_{\delta\gamma}s^{a}_{\kappa\lambda}\right) 
\left( \sigma^{b}_{\gamma\delta}s^{b}_{\nu\mu}\right) 
\left( \sigma^{c}_{\alpha\beta}s^{c}_{\mu\nu}\right)\nonumber\\ 
\label{eq:gamma_spin-d}\\
\hat\Gamma^{(x)}_{\alpha\beta, \kappa\lambda} &=& \hspace*{-0.2cm}
\sum_{a,b,c=x,y,z}\ \sum_{\gamma,\delta,\mu,\nu=1}^2
\left( \sigma^{a}_{\delta\gamma}s^{a}_{\kappa\lambda}\right) 
\left( \sigma^{b}_{\alpha\delta}s^{b}_{\nu\mu}\right) 
\left( \sigma^{c}_{\gamma\beta}s^{c}_{\mu\nu}\right)\nonumber\\ 
\label{eq:gamma_spin-x}
\end{eqnarray}
with $c-$electron spin indices $\alpha$, $\beta$, $\gamma$, $\delta$, and 
$f-$spin indices $\kappa$, $\lambda$, $\mu$, $\nu$, 
as shown in Fig.~\ref{fig:RKKYvertex}~(a). The spin summations can be 
performed using the spin algebra ($a,b=x,y,z$), 
\begin{eqnarray}
\sum_{\gamma=1}^2
\sigma^{a}_{\alpha\gamma}\sigma^{b}_{\gamma\beta} = \sum_{c=x,y,z}
i \varepsilon^{abc} \sigma^{c}_{\alpha\beta}
+\delta^{ab}\mathds{1}_{\alpha\beta} \ ,
\label{eq:spin_algebra}
\end{eqnarray}
where $\mathds{1}$ is the unit operator in spin space,   
$\varepsilon^{abc}$ the totally antisymmetric tensor and 
$\delta^{ab}$ the Kronecker-$\delta$. This results in a nonlocal 
Heisenberg coupling between sites $i$ and $j$,   
\begin{eqnarray}
\hat\Gamma^{(d)}_{\alpha\beta, \kappa\lambda} &=& 
\phantom{+}4 \sum_{a=x,y,z}
\left( \sigma^{a}_{\alpha\beta}s^{a}_{\kappa\lambda}\right) 
\label{eq:gamma_spin_d2}\\
\hat\Gamma^{(x)}_{\alpha\beta, \kappa\lambda} &=& 
-2 \sum_{a=x,y,z}
\left( \sigma^{a}_{\alpha\beta}s^{a}_{\kappa\lambda}\right) 
\label{eq:gamma_spin_x2} \ .
\end{eqnarray}

\subsection{Energy dependence}
\label{subsec:RKKYvertex-energy}

With the energy variables as defined in Fig.~\ref{fig:RKKYvertex}~(b),
the energy dependent functions in \Eq{eq:RKKYvertex} read 
in Matsubara representation 
\begin{eqnarray}
\Lambda^{(d)}_{RKKY}({\bf r}_{ij},i\Omega_m) &=& 
JJ_0^2\Pi({\bf r}_{ij},i\Omega_m)\tilde\chi_f(i\Omega_m) 
\nonumber
\label{eq:Lambda_d0}\\ 
\Lambda^{(x)}_{RKKY}({\bf r}_{ij},i\omega_n,i\Omega_m) &=& 
-JJ_0^2 T\sum_{i\varepsilon_m} 
G_c({\bf r}_{ij},i\omega_n+i\varepsilon_m)
G_c({\bf r}_{ij},i\omega_n+i\Omega_m+i\varepsilon_m)\tilde\chi_f(i\varepsilon_m)
\nonumber
\label{eq:Lambda_x0}
\end{eqnarray}
where
\begin{eqnarray}
\Pi({\bf r}_{ij},i\Omega_m) &=& - T\sum_{\varepsilon_n}
G_c({\bf r}_{ij},i\varepsilon_n)G_c({\bf r}_{ij},i\varepsilon_n+i\Omega_m) \hspace*{2.1cm}
\label{eq:chic}
\end{eqnarray}
and $\tilde\chi_f(i\varepsilon_m)=\chi_f(i\varepsilon_m)/(g_L\mu_B)^2$, 
with $\chi_f(i\varepsilon_m)$ the full, single-impurity $f$-spin 
susceptibility, \Eq{eq:chif_ret}.

For the renormalization of the total $c-f$ vertex for $c-$electrons 
at the Fermi energy, 
the contributions $\Lambda^{(d)}_{RKKY}$, $\Lambda^{(d)}_{RKKY}$
must be calculated for real frequencies, $i\Omega\to \Omega+i0$,
$i\omega\to \omega+i0$, and for electrons at the Fermi energy, i.e.,
$\omega=0$. In this limit, only the real parts of 
$\Lambda^{(d)}_{RKKY}$, $\Lambda^{(d)}_{RKKY}$ contribute to the vertex 
renormalization, as seen below.
In order to analyze their importance for the RG flow, we will
expand them in terms of the small parameter $T_K/D_0$.
In the following, the real part of a complex function will be denoted 
by a prime {'} and the imaginary part by a double-prime {''}.\\[0.2cm]

{\bf Direct contribution.}
Since in  $\Lambda^{(d)}_{RKKY}$, $\Pi(i\Omega_m)$ and 
$\tilde\chi_f(i\Omega_m)$ appear as a product and $\tilde\chi_f(\Omega)$ 
cuts off the energy transfer $\Omega$ at the 
scale $T_K\ll \varepsilon_F\approx D_0$, the electron polarization 
$\Pi(\Omega)$ contributes only in the limit $\Omega \ll \varepsilon_F$ 
where it is real-valued, as seen in \Eq{eq:chic_ret}. 
Using \Eq{eq:chic_ret} and \Eq{eq:chif_ret}, the real part of the 
direct RKKY-induced vertex contribution reads,
\begin{eqnarray}
\Lambda^{(d)}_{RKKY}{'}({\bf r}_{ij},\Omega+i0) &=& 
JJ_0^2  R({\bf r}_{ij}) A N(0) \, 
\frac{D_0}{T_K}\,\frac{1}{\sqrt{1+(\Omega/T_K)^2}} \, +\,
{\cal O}\left(\left(\frac{\Omega}{D_0}\right)^2\right),
\label{eq:Lambda_d1}
\end{eqnarray}
where   
\begin{eqnarray}   
R({\bf r}_{ij}) &=& \frac{\sin(x)-x\cos(x)}{4x^4}, \qquad\qquad x=2k_Fr
\end{eqnarray}   
is a spatially oscillating function.\\[0.3cm]
 
{\bf Exchange contribution.}
In order to analyze the size of $\Lambda^{(x)}_{RKKY}{'}$ in terms of 
$T_K/D_0$, it is sufficient to evaluate it for a particle-hole symmetric 
conduction band and for ${\bf r}_{ij}=0$, since  
the $T_K/D_0$ dependence is induced by the on-site susceptibility
$\tilde\chi_f(i\Omega)$. The dependence on $T_K/D_0$ can be changed by 
the frequency convolution involved in $\Lambda^{(x)}_{RKKY}{'}$, but does not 
depend on details of the conduction band and distance dependent 
terms. (The general calculation is possible as well, but considerably 
more lengthy.)    
We use the short-hand notation for the momentum-integrated
$c-$electron Green's function,
$G_c({\bf r}=0,\omega \pm i0)=G(\omega)=G'(\omega)+iG''(\omega)$, and
assume a flat density of states $N(\omega)$, with the upper and lower
band cutoff symmetric about $\varepsilon_F$, i.e., 
\begin{eqnarray}
G^{R/A}{''}(\omega) &=& \mp \frac{\pi}{2D_0}\Theta(D_0-|\omega|)\\
G^{R/A}{'}(\omega)  &=& \frac{1}{2D_0} 
\ln\left|\frac{D_0+\omega}{D_0-\omega}\right| = \frac{\omega}{D_0^2} 
\,+\,{\cal O}\left( \left(\frac{\omega}{D_0}\right) \right) \ .
\end{eqnarray}   
Furthermore, at $T=0$ the Fermi and Bose distribution functions are,
$f(\varepsilon)=-b(\varepsilon)=\Theta(-\varepsilon)$.\\
$\Lambda^{(x)}_{RKKY}{'}(0,0,\Omega+i0)$ then reads, 
\begin{eqnarray}
\Lambda^{(x)}_{RKKY}{'}({\bf r}_{ij}=0,\omega=0+i0,\Omega+i0) &=& \nonumber \\ 
&& \hspace*{-7.3cm}- JJ_0^2 \, \left\{  
\,\int \frac{d\varepsilon}{\pi}
\left[
f(\varepsilon) G^A{''}(\varepsilon)G^R{'}(\varepsilon+\Omega) 
+ f(\varepsilon+\Omega)\, G^A{'}(\varepsilon)G^A{''}(\varepsilon+\Omega)
\right] \, \tilde\chi_f^R{'}(\varepsilon) \right.
 \label{eq:Lambda_x}\\
&& \hspace*{-6.25cm}
\left. -\,\int \frac{d\varepsilon}{\pi}
\left[
f(\varepsilon)\, G^R{'}(\varepsilon)G^R{'}(\varepsilon+\Omega) 
- f(\varepsilon+\Omega) G^A{''}(\varepsilon)G^A{''}(\varepsilon+\Omega)
\right] \, \tilde\chi_f^R{''}(\varepsilon) \right\}
\nonumber \ .
\end{eqnarray}
With the above definitions, the four terms in this expression are evaluated 
in an elementary way, using the substitution $\varepsilon_F/T_K=x=\sinh u$,
\begin{eqnarray}
\int \frac{d\varepsilon}{\pi}
f(\varepsilon) G^A{''}(\varepsilon)G^R{'}(\varepsilon+\Omega) 
\tilde\chi_f^R{'}(\varepsilon) && \nonumber \\ 
&&\hspace*{-4.0cm} =A N(0) \frac{T_K}{D_0}\, \left[
1-\sqrt{1+\left(\frac{D_0}{T_K}\right)^2}+
\frac{\Omega}{T_K}\, {\rm arsinh}\left(\frac{D_0}{T_K}\right)
\right] \nonumber\\
&&\hspace*{-4.0cm} =A N(0) \left[ -1 + 
\frac{\Omega}{D_0}\,  \ln\left(\frac{D_0}{T_K}\right)
\, + \, 
{\cal O}\left(\frac{T_K}{D_0}\right)
\right]
\label{eq:Lambda_x1}
\end{eqnarray}
\begin{eqnarray}
\left|\int \frac{d\varepsilon}{\pi}
f(\varepsilon+\Omega)\, G^A{'}(\varepsilon)G^A{''}(\varepsilon+\Omega)
\tilde\chi_f^R{'}(\varepsilon) \right| && \nonumber \\
&&\hspace*{-4.5cm}=A N(0) \frac{T_K}{D_0} 
\left|
\sqrt{1+\left(\frac{\Omega}{T_K}\right)^2} -
\sqrt{1+\left(\frac{D_0+\Omega}{T_K}\right)^2} 
\right| \nonumber\\
&&\hspace*{-4.5cm}\leq A N(0) \,+\, {\cal O}\left(\frac{T_K}{D_0}\right)
\label{eq:Lambda_x2}
\end{eqnarray}   

\begin{eqnarray}
\int \frac{d\varepsilon}{\pi}
f(\varepsilon)\, G^R{'}(\varepsilon)G^R{'}(\varepsilon+\Omega) 
\tilde\chi_f^R{''}(\varepsilon) &&\nonumber\\
&&\hspace*{-4.0cm}
=-\frac{4}{\pi^2} A N(0) \left(\frac{1}{2}+\frac{\Omega}{D_0} \right)\,
\ln\left(\frac{D_0}{T_K}\right) 
\,+\, {\cal O}\left(\left(\frac{T_K}{D_0}\right)^0\right)
\label{eq:Lambda_x3}
\end{eqnarray}   

\begin{eqnarray}
\int \frac{d\varepsilon}{\pi}
f(\varepsilon+\Omega) G^A{''}(\varepsilon)G^A{''}(\varepsilon+\Omega)
\tilde\chi_f^R{''}(\varepsilon) && \nonumber\\
&&\hspace*{-4.5cm}=\frac{\pi}{4}A N(0) \left[
- {\rm arsinh}\left(\frac{\Omega}{T_K}\right) 
+ {\rm arsinh}\left({\rm min}
\left(\frac{\Omega}{T_K},\frac{D_0+\Omega}{T_K}\right)\right) 
\right]\nonumber \\
&&\hspace*{-4.5cm}\leq \frac{\pi}{4}A N(0) 
\,+\, {\cal O}\left(\frac{T_K}{D_0}\right) \ .
\label{eq:Lambda_x4}
\end{eqnarray}   

Comparing  Eqs.~(\ref{eq:Lambda_x})--(\ref{eq:Lambda_x4}) with
\Eq{eq:Lambda_d1} shows that all terms of $\Lambda^{(x)}_{RKKY}{'}(\Omega)$ 
are subleading compared to $\Lambda^{(d)}_{RKKY}{'}(\Omega)$ by at least 
a factor $(T_K/D_0)\ln(T_K/D_0)$ for all transfered energies $\Omega$.
Hence, it can be neglected in the RG flow. 
Combining the results of spin and energy dependence, 
Eqs.~(\ref{eq:RKKYvertex}), (\ref{eq:gamma_spin_d2}) and (\ref{eq:Lambda_d1}), 
one obtains the total RKKY-induced $c-f$ vertex as, 
\begin{eqnarray} 
\hat\gamma_{RKKY}^{(d)} ({\bf r}_{ij},i\Omega)&=&
2 \,(1-\delta_{ij})\,\Pi({\bf r}_{ij},i\Omega)\chi_f(i\Omega)\,{\bf S}_i
\cdot {\bf s}_j
\end{eqnarray}
or
\begin{eqnarray} 
{\rm Re}\hat\gamma_{RKKY}^{(d)} ({\bf r}_{ij},\Omega+i0)
\hspace*{-0.2cm}&=&\hspace*{-0.2cm}
2 JJ_0^2 AN(0)\,(1-\delta_{ij}) R({\bf r}_{ij}) \,
\frac{D_0}{T_K}\,\frac{1}{\sqrt{1+(\Omega/T_K)^2}} 
\,{\bf S}_i \cdot {\bf s}_j .
\end{eqnarray}


\clearpage
%

\bibliographystyle{correl}
\bibliography{LectureNotes_Kondo-RKKY_Kroha}

\begin{thebibliography}{10}
\providecommand{\url}[1]{\texttt{#1}}
\providecommand{\urlprefix}{}
\expandafter\ifx\csname urlstyle\endcsname\relax
  \providecommand{\doi}[1]{doi:\discretionary{}{}{}#1}\else
  \providecommand{\doi}{doi:\discretionary{}{}{}\begingroup
  \urlstyle{rm}\Url}\fi

\bibitem{Kondo64}
J.~Kondo, Prog. Theor. Phys \textbf{32}, 37 (1964)

\bibitem{Hewson93}
A.~C. Hewson: \emph{The Kondo Problem to Heavy Fermions} (Cambridge University
  Press, 1993)

\bibitem{Ruderman54}
M.~A. Ruderman and C.~Kittel, Phys. Rev. \textbf{96}, 99 (1954)

\bibitem{Kasuya56}
T.~Kasuya, Prog. Theor. Phys. \textbf{16}, 45 (1956)

\bibitem{Yosida57}
K.~Yosida, Phys. Rev. \textbf{106}, 893 (1957)

\bibitem{Loehneysen07}
H.~v. L\"ohneysen, A.~Rosch, M.~Vojta, and P.~W\"olfle, Rev. Mod. Phys.
  \textbf{79}, 1015 (2007)

\bibitem{Doniach77}
S.~Doniach, Physica B+C \textbf{91}, 231  (1977)

\bibitem{Hertz76}
J.~A. Hertz, Phys. Rev. B \textbf{14}, 1165 (1976)

\bibitem{Moriya85}
T.~Moriya: \emph{Spin fluctuations in itinerant electron magnetism} (Springer,
  Berlin, 1985)

\bibitem{Millis93}
A.~J. Millis, Phys. Rev. B \textbf{48}, 7183 (1993)

\bibitem{Si01}
K.~Q.~Si, S.~Rabello and J.~L. Smith, Nature \textbf{413}, 804 (2001)

\bibitem{Coleman01}
P.~Coleman, C.~P\'epin, Q.~Si, and R.~Ramazashvili, Journal of Physics:
  Condensed Matter \textbf{13}, R723 (2001)

\bibitem{Senthil04}
T.~Senthil, M.~Vojta, and S.~Sachdev, Phys. Rev. B \textbf{69}, 035111 (2004)

\bibitem{Woelfle11}
P.~W\"olfle and E.~Abrahams, Phys. Rev. B \textbf{84}, 041101 (2011)

\bibitem{Woelfle14}
E.~Abrahams, J.~Schmalian, and P.~W\"olfle, Phys. Rev. B \textbf{90}, 045105
  (2014)

\bibitem{Woelfle16}
P.~W\"olfle and E.~Abrahams, Phys. Rev. B \textbf{93}, 075128 (2016)

\bibitem{Abrikosov65}
A.~A. Abrikosov, Physics \textbf{2}, 21 (1965)

\bibitem{Barnes76}
S.~E. Barnes, J. Phys. F \textbf{6}, 1375 (1976)

\bibitem{Coleman84}
P.~Coleman, Phys. Rev. B \textbf{29}, 3035 (1984)

\bibitem{Kroha98}
J.~Kroha and P.~W{\"o}lfle, Acta Phys. Pol. B \textbf{29}, 3781 (1998)

\bibitem{Kroha97}
J.~Kroha, P.~W{\"o}lfle, and T.~A. Costi, Phys. Rev. Lett. \textbf{79}, 216
  (1997)

\bibitem{Andrei83}
N.~Andrei, K.~Furuya, and J.~H. Lowenstein, Rev. Mod. Phys. \textbf{55}, 331
  (1983)

\bibitem{Bulla08}
R.~Bulla, T.~A. Costi, and T.~Pruschke, Rev. Mod. Phys. \textbf{80}, 395 (2008)

\bibitem{Anderson70}
P.~W. Anderson, Journal of Physics C: Solid State Physics \textbf{3}, 2436
  (1970)
\newline\urlprefix\url{http://stacks.iop.org/0022-3719/3/i=12/a=008}

\bibitem{Jones88}
B.~A. Jones, C.~M. Varma, and J.~W. Wilkins, Phys. Rev. Lett. \textbf{61}, 125
  (1988)

\bibitem{Affleck95}
I.~Affleck, A.~W.~W. Ludwig, and B.~A. Jones, Phys. Rev. B \textbf{52}, 9528
  (1995)

\bibitem{Nejati17}
A.~Nejati, K.~Ballmann, and J.~Kroha, Phys. Rev. Lett. \textbf{118}, 117204
  (2017)

\bibitem{Lambert}
D.~Veberi\'c, Computer Phys. Commun. \textbf{183}, 2622 (2012)
\newline\urlprefix\url{arXiv:1209.0735}

\bibitem{Bork11}
J.~Bork, Y.-H. Zhang, L.~Diekh\"oner, L.~Borda, P.~Simon, J.~Kroha, P.~Wahl,
  and K.~Kern, Nat. Phys. \textbf{7}, 901 (2011)

\bibitem{Kroeger11}
N.~N\'eel, R.~Berndt, J.~Kr\"oger, T.~O. Wehling, A.~I. Lichtenstein, and M.~I.
  Katsnelson, Phys. Rev. Lett. \textbf{107}, 106804 (2011)

\end{thebibliography}

\clearchapter


\end{document}